\documentclass[fleqn,usenatbib]{mnras}

\usepackage{newtxtext,newtxmath}

\usepackage[T1]{fontenc}

\DeclareRobustCommand{\VAN}[3]{#2}
\let\VANthebibliography\thebibliography
\def\thebibliography{\DeclareRobustCommand{\VAN}[3]{##3}\VANthebibliography}

\usepackage{graphicx}	
\usepackage{amsmath}	
\usepackage{ulem}
\usepackage{multirow}
\usepackage{pstool}
\usepackage{psfrag}
\usepackage{pdflscape}

\newcommand{\V}{\mathcal{V}}
\newcommand{\cl}{$C_{\ell}(\Delta\nu)$}
\newcommand{\pk}{$P(k_{\perp}, k_{\parallel})$}
\newcommand{\pkb}{$P(\mathbf{k_{\perp}}, k_{\parallel})$}
\newcommand{\maps}{$C_{\ell}(\nu_a,\nu_b)$}
\newcommand{\HI}{\ion{H}{i}}
\newcommand{\omb}{$[\Omega_{\HI} b_{\HI}]$}
\newcommand{\ombsq}{$[\Omega_{\HI} b_{\HI}]^2$}
\newcommand{\ellb}{\boldsymbol{\ell}}
\newcommand{\kb}{\boldsymbol{k_{\perp}}}
\newcommand{\clb}{$C_{\ellb}(\Delta\nu)$}
\newcommand{\mapsb}{$C_{\ellb}(\nu_a,\nu_b)$}

\newcommand{\abyda}{\mid \mathbf{A}\mid / \mathbf{dA}}

\defcitealias{Ch21}{Ch21} 
\defcitealias{P22}{Paper~I}
\defcitealias{AE23}{Paper~II}
\defcitealias{AE23b}{Paper~III} 

\title[Wideband intensity mapping]{Towards $21$-cm intensity mapping at $z=2.28$ with uGMRT using the tapered gridded estimator -- IV. Wideband analysis}

\author[A. Elahi et al.]{Khandakar Md Asif Elahi,$^{1}$\thanks{E-mail:asifelahi999@gmail.com}
Somnath Bharadwaj,$^{1}$\thanks{E-mail:somnath@phy.iitkgp.ac.in}
Srijita Pal,$^{2}$
Abhik Ghosh,$^{3}$
Sk. Saiyad Ali,$^{4}$
\newauthor
Samir Choudhuri,$^{5}$
Arnab Chakraborty,$^{6}$
Abhirup Datta,$^{7}$
Nirupam Roy,$^{2}$
Madhurima Choudhury$^{8}$ 
\newauthor
and Prasun Dutta$^{9}$
\\
\\
$^{1}$ Department of Physics and Centre for Theoretical Studies, IIT Kharagpur, Kharagpur 721 302, India\\
$^{2}$ Department of Physics, Indian Institute of Science, Bangalore 560012, India\\
$^{3}$ Department of Physics, Banwarilal Bhalotia College, Asansol, West Bengal-713303, India\\
$^{4}$ Department of Physics, Jadavpur University, Kolkata 700032, India\\
$^{5}$ Centre for Strings, Gravitation and Cosmology, Department of Physics, Indian Institute of Technology Madras, Chennai 600036, India\\
$^{6}$ Department of Physics and McGill Space Institute, McGill University, Montreal, QC, Canada H3A 2T8\\
$^{7}$ Discipline of Astronomy, Astrophysics and Space Engineering, Indian Institute of Technology Indore, Indore 453552, India\\
$^{8}$ ARCO (Astrophysics Research Center), Department of Natural Sciences, The Open University of Israel, 1 University Road, PO Box 808, Ra’anana 4353701, Israel\\
$^{9}$ Department of Physics, IIT (BHU), Varanasi, 221005 India
}

\date{Accepted XXX. Received YYY; in original form ZZZ}

\pubyear{2023}

\begin{document}
\label{firstpage}
\pagerange{\pageref{firstpage}--\pageref{lastpage}}
\maketitle

\begin{abstract}
     We present a Wideband Tapered Gridded Estimator (TGE), which incorporates baseline migration and variation of the primary beam pattern for neutral hydrogen ($\ion{H}{i}$) 21-cm intensity mapping (IM) with large frequency bandwidth radio-interferometric observations. Here we have analysed  $394-494 \, {\rm MHz}$  $(z = 1.9 - 2.6)$  uGMRT data to estimate the Multi-frequency Angular Power Spectrum (MAPS) $C_\ell(\Delta\nu)$ from which we have removed the foregrounds using the polynomial fitting (PF) and Gaussian Process Regression (GPR) methods developed in our earlier work.  Using the residual  $C_\ell(\Delta\nu)$  to estimate the mean squared 21-cm brightness temperature fluctuation $\Delta^2(k)$, we find that this is consistent with $0 \pm 2 \sigma$ in several $k$ bins.  The resulting $2\sigma$ upper limit $\Delta^2(k) < (4.68)^2 \, \rm{mK^2}$ at $k=0.219\,\rm{Mpc^{-1}}$  is nearly $15$ times tighter than earlier limits obtained from a smaller bandwidth ($24.4 \, {\rm MHz}$)  of the same data. The $2\sigma$ upper limit $[\Omega_{\ion{H}{i}} b_{\ion{H}{i}}] < 1.01 \times 10^{-2}$ is within an order of magnitude of the value expected from independent estimates of the $\ion{H}{i}$ mass density $\Omega_{\ion{H}{i}}$ and the $\ion{H}{i}$ bias $b_{\ion{H}{i}}$. The techniques used here can be applied to other telescopes and frequencies, including $\sim 150 \, {\rm MHz}$ Epoch of Reionization observations. 
\end{abstract}




\section{Introduction}

The redshifted $21$-cm line  from the cosmological distribution 
of neutral hydrogen (\HI{}) atoms appears as a faint background radiation at frequencies below  $1420\,\rm{MHz}$.  Mapping the spectral and angular fluctuations in the specific intensity of this  radiation  (Intensity Mapping - IM) holds the potential to quantify the large-scale clustering of the matter distribution  \citep{BNS}. The first detection of this signal by 
\cite{Pen2009a}  cross-correlated the $z\sim0.02$ IM signal from the \HI{} Parkes All Sky Survey  (\citealt{Barnes2001})  with the optical six-degree field galaxy redshift survey (\citealt{Jones2004, Jones2005}). \cite{chang10} reported a $4\sigma$  detection at $z=0.8$ cross-correlating the Green Bank Telescope (GBT) radio observation and optical DEEP2 Redshift Survey \citep{Deep2}. 
Several subsequent works have reported the detection of the IM signal in cross-correlation at $z\sim 1$ using both single-dish telescopes \citep{masui2013, SW13, Anderson2018, Wolz2022, Cunnington23} and interferometers \citep{chime22}. \cite{CHIMELya} recently detected the signal in the redshift range $1.8<z<2.5$ cross-correlating $1000$ hours data from the Canadian Hydrogen Intensity Mapping Experiment (CHIME) with the Lyman-$\alpha$ forest measurements from the Extended Baryon Oscillation Spectroscopic Survey (eBOSS; \citealt{Bourboux2020}).
\cite{ghosh1,ghosh2} have attempted to detect the IM  signal in auto-correlation with Giant Metrewave Radio Telescope (GMRT; \citealt{swarup91}) data at $z=1.32$, and have obtained an upper limit $[\Omega_{\HI}b_{\HI}] < 0.11$ where $\Omega_{\HI}$ and $b_{\HI}$ are the \HI{} mass density and the \HI{} bias parameters, respectively. \cite{Paul23} recently reported the first  detection of the 21-cm signal from $z=0.32$ and $0.44$ in auto-correlation using $96$ hours of observation with the MeerKAT interferometer.      

The recently upgraded GMRT (uGMRT; \citealt{uGMRT}) with the GMRT Wideband Backend (GWB) provides a large instantaneous frequency coverage, which opens up the possibility for IM studies over a large redshift range.  A 25-hour observation of the European Large-Area ISO Survey-North 1 (ELAIS-N1) field was performed using the Band~3 (300--500~MHz)  of   uGMRT in May 2017. \cite{Cha2} first identified and removed the point sources (having flux densities $> 100 \,\mu {\rm Jy}$) from this data, and used the residual data to characterize the diffused Galactic foregrounds for the entire 200~MHz bandwidth. \cite{Ch21} (hereafter \citetalias{Ch21}) further estimated the 21-cm power spectrum (PS) using the delay space technique. Frequency channels which are flagged to avoid human-made Radio Frequency Interference (RFI) introduce artefacts in  delay space, which \citetalias{Ch21} compensated for using one-dimensional \textsc{clean} \citep{Parsons_2009}. \citetalias{Ch21} further used foreground avoidance to obtain upper limits $[\Omega_{\HI}b_{\HI}] \lesssim 0.1 - 0.24$ for several redshifts in the  range  $1.96 < z < 3.58$ at $k\sim 1 \, {\rm Mpc^{-1}}$. 

The present paper is the fourth in a series of works which have used the uGMRT observation mentioned in the previous paragraph.  In these works, we have used the Tapered Gridded Estimator (TGE; \citealt{samir14, samir16, samir17}) which offers several unique features for estimating the 21-cm PS from radio-interferometric measurements.  TGE tapers the sky response of the telescope to reduce the contribution from wide-field foregrounds like bright extra-galactic sources located in the side-lobes.  On a different note, missing frequency channels pose a significant difficulty in 21-cm PS estimation, and several algorithms have been proposed to deal with this  \citep{Parsons_2009, Trott2016, Ewall-Wice2021,  Kern2021, Kennedy2023}.   TGE naturally overcomes this difficulty as it  first estimates  \cl{} the  Multi-frequency Angular Power Spectrum (MAPS). Typically, we get the  estimated \cl{} for all frequency separation $\Delta \nu$,  even if several frequency channels are missing. The 21-cm PS, which  is estimated by Fourier transforming \cl{} with respect $\Delta \nu$, is found to be free of artefacts \citep{Bh18}. \cite{Pal20} have demonstrated the potentials of TGE by applying it on 150~MHz GMRT data where $\sim 50 \%$ of the data are flagged. 

In \cite{P22}, which is  \citetalias{P22} of this series,  we have used TGE to analyse a $24.4\,\rm{MHz}$ sub-band of the uGMRT data mentioned earlier.  This sub-band has the central frequency of $432.8 \, \rm {MHz}$  $(z=2.28)$.  The signal in the two  polarization states  (RR and LL) were combined for the PS  estimation and an upper limit of $[\Omega_{\HI}b_{\HI}] < 0.23$ 
at $k = 0.347 \, {\rm Mpc}^{-1}$ was obtained.  Continuing with the same data, in \cite{AE23} (\citetalias{AE23}) we showed that the foreground level is substantially reduced if the PS is estimated by cross-correlating  the RR and LL polarizations rather than combining them, and the upper limit was tightened to 
$[\Omega_{\HI}b_{\HI}] < 6.1 \times 10^{-2}$ at $k = 0.804 \, {\rm Mpc^{-1}}$.  Both the above works found a considerable part of the $(k_{\perp},k_{\parallel})$ plane to be foreground-contaminated, and it was necessary to avoid this region to estimate the 21-cm PS. 
In  \cite{AE23b} (\citetalias{AE23b}) we have introduced a foreground removal technique whereupon we are able to utilise the whole  $(k_{\perp},k_{\parallel})$ plane to estimate the 21-cm PS and obtain an even tighter upper limit  $[\Omega_{\HI}b_{\HI}] < 2.2 \times 10^{-2}$ at a larger scale of $k = 0.247 \, {\rm Mpc^{-1}}$ as compared to the earlier works. 

The earlier works with the uGMRT data discussed above have all been restricted to small sub-bands ($< 25 \, {\rm MHz})$ of the entire Band~3 frequency coverage. While it is desirable to consider the largest possible bandwidths so as to increase the signal-to-noise ratio (SNR) and also access smaller $k_{\parallel}$ modes, it then becomes imperative to account for the fact that the properties of the instrument changes with frequency.  Foremost among these is `baseline migration' {\it i.e.} the baseline corresponding to a fixed antenna pair changes with frequency.  Further, the primary beam (PB)   of the individual antennas also changes with frequency,  a fact which affects the normalisation of the estimated 21-cm PS. It is acceptable to ignore these effects over small observational bandwidths, as is the case for many of the visibility-based estimators used for the  EoR 21-cm PS (e.g., \citealt{parsons12, mertens18}).  We note that the earlier implementation of TGE does not incorporate baseline migration, although it does account for the frequency-dependent variation of the PB. In the present paper we introduce the Wideband TGE  (hereafter referred to as the TGE) which also accounts for baseline migration, and is thus well suited for analysing large frequency bandwidths.  

\begin{figure}
    \centering
    \includegraphics[width=\columnwidth, bb=134.7 271.8 545.1 514.5]{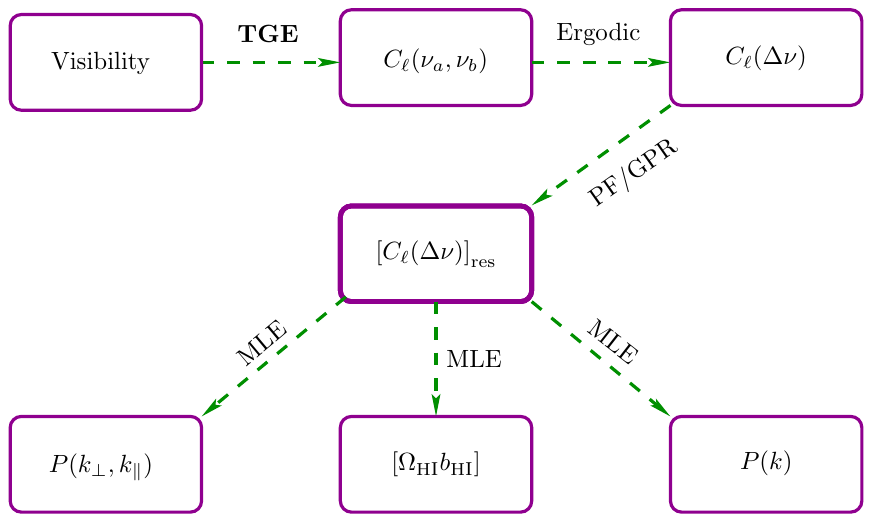}
    \caption{{ This shows the workflow of this paper. } }
    \label{fig:tge_diagram}
\end{figure} 

In this paper, we present the analysis of 100~MHz bandwidth data which span the frequency range $394-494\,{\rm MHz}$ corresponding to the redshift band $z=1.9-2.6$. Although the TGE can now be used to analyse the entire $200 \, {\rm MHz}$ Band~3 coverage, we have discarded the rest of the data due to their poor quality. The data is described in Section~\ref{sec:data}. Figure~\ref{fig:tge_diagram} presents the workflow we have undertaken for analysing the data. We have estimated the MAPS \mapsb{} by applying the TGE on the visibility data. The TGE is described in \citetalias{AE23b}, and in Section~\ref{sec:wtge} of the present paper we have highlighted the changes we have incorporated in TGE for the wideband IM. For the subsequent analysis, we have assumed that the 21-cm signal is ergodic (statistically homogeneous) along the line-of-sight direction. This  allows us to consider the MAPS \clb{} instead of \mapsb{}, where $C_{\ellb} (\Delta\nu) = C_{\ellb} (\nu_a, \nu_b)  = C_{\ellb} (\mid \nu_a - \nu_b \mid)$. Considering the post-reionization \HI{} 21-cm signal,  such an assumption is justified because various observations \citep{Prochaska2005, Noterdaeme09, Not, Zafar, Bird17} and simulations \citep{Deb16} respectively indicate that  $\Omega_{\HI{}}(z)$ and $[b_{\HI{}}(z) \, D(z)]$  (where $D(z)$ is the growing mode of linear density perturbations) are nearly constant for our observational redshift interval.

We remove the foregrounds from \clb{} using the methodologies developed in \citetalias{AE23b}. The methods are briefly described in Section~\ref{sec:FGremoval} and \citetalias{AE23b} is referred to for the details.  After foreground removal, we use the residual $\left[ C_\ell(\Delta\nu) \right]_{\rm res}$ to obtain cylindrical PS $P(k_\perp, k_\parallel)$, spherical PS $P(k)$ and the cosmological \HI{} abundance \omb{} which are described in Sections~\ref{sec:cylindricalps}, ~\ref{sec:BMLE} and ~\ref{sec:omb}, respectively. We summarize our results and takeaways from this analysis in Section~\ref{sec:conclusion}. 

The paper uses the same values of the cosmological parameters used in our earlier works, where the values are taken from \cite{Planck18f}.

\section{Data Description}
\label{sec:data}

\begin{figure*}
    \centering
    \includegraphics[width=\textwidth]{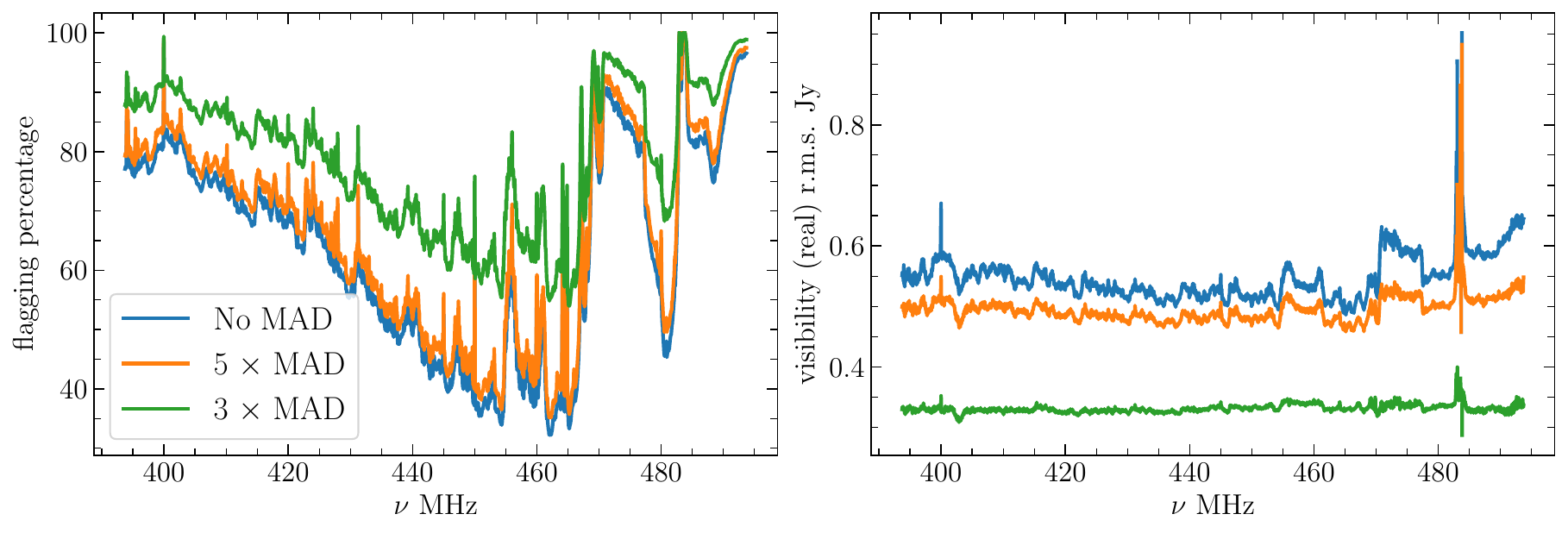}
    \caption{This shows the flagging percentage (left) and the r.m.s. of the visibilities (right) as a function of frequency. The blue, orange, and green curves show the cases with No MAD filtering, $5\times \rm{MAD}$ filtering, and $3\times \rm{MAD}$ filtering, respectively. }
    \label{fig:flagrms}
\end{figure*}
\begin{figure*}
    \centering
    \includegraphics[width=\textwidth]{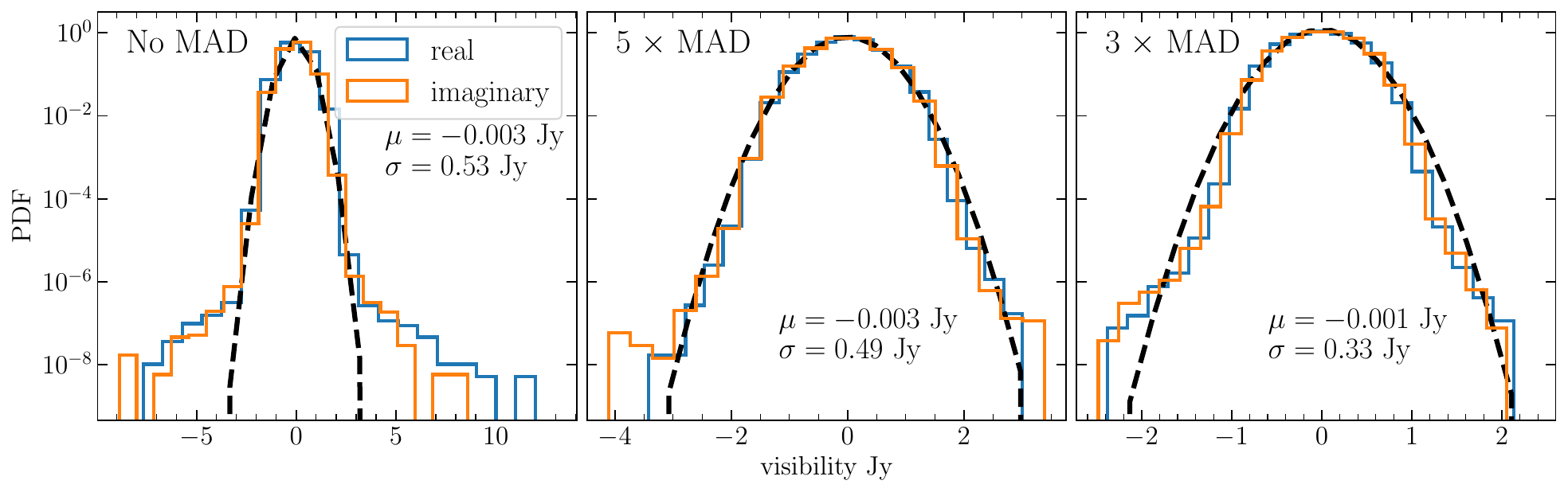}
    \caption{The left panel shows the probability density function (PDF) of the real and imaginary parts of the visibility data. The black dashed lines represent a fit of the PDF of the real part of the visibilities with a Gaussian distribution whose mean and standard deviation are quoted in the figure. The cases with $5\times \rm{MAD}$ and $3\times \rm{MAD}$ filtering are shown in the middle and the right panels, respectively.}
    \label{fig:madstat}
\end{figure*}

The details of the observation and initial data pre-processing are described in \cite{Cha2}. A brief summary of the major steps can be found in our earlier works (\citetalias{P22}, \citetalias{AE23} and \citetalias{AE23b}).
In short, the  observation spans $1.8 \,\rm{deg}^2$ on the sky, and the visibility data have 200~MHz bandwidth $(300 - 500 \,{\rm MHz})$ with the resolution $\Delta \nu_c = 24.4\,\rm{kHz}$. A direction-independent calibration has been applied to the data without doing any polarization calibration. Here, we use the residual data where bright sources (flux densities $> 100\, \mu {\rm Jy}$) are identified and removed using the \textsc{uvsub} routine of the Common Astronomy Software Application (\textsc{casa}; \citealt{casa07}).  


In the initial data analysis \citep{Cha2},  5~MHz from the edges of the bandwidth and $90\%$ of the data in the frequency range $300 - 390 \,{\rm MHz}$ are flagged due to RFIs using the software \textsc{aoflagger} \citep{Off10, Off12} and the \textsc{rflag} routine of \textsc{casa}. We have entirely discarded these in this work. Considering the rest, we have split out $100$ MHz bandwidth data in the frequency range $394-494\,\rm{MHz}$ for the present analysis. Initially the data had $\sim64\%$ flagging and a visibility r.m.s. of $0.54\,\rm{Jy}$. The blue curves in Figure~\ref{fig:flagrms} show the flagged percentage (left) and the r.m.s. of the real part of the visibility data (right). We see that the flagged percentage varies significantly across the frequency band, where it is $<60\%$ in the frequency range $430-470\,\rm{MHz}$ and $\approx80 \%$ at the two ends of the band. As for the r.m.s., we see that it is nearly constant $(0.54\,{\rm Jy})$ across the band except for the frequency range $470-494$~MHz, where the r.m.s. is somewhat  higher $(0.6 - 0.8 \,{\rm Jy})$.

The leftmost panel of Figure~\ref{fig:madstat} shows the probability density function (PDF) of the real and imaginary parts of the visibility data. We see that the central part of the PDF matches with a Gaussian distribution of mean $\mu\approx0\,{\rm Jy}$ and standard deviation $\sigma\approx0.53\, {\rm Jy}$. However, the PDF has large positive and negative values which are not  predicted by the Gaussian distribution. We have identified and flagged these large outlier values by considering the median absolute deviation (MAD), which for a given data is defined as the median of the absolute deviations from the median of the data\footnote{If $M$ is the median of the array $(x_1, x_2, x_3, .......)$, then MAD is the median of the array $(\mid x_1 - M \mid, \mid x_2 - M\mid, \mid x_3 - M\mid, ...)$} (see, e.g. \citealt{jayanti}). For each baseline, we have calculated the MAD  and flagged the frequency channels where the visibility amplitude had values significantly larger than the MAD value. The middle panel of Figure~\ref{fig:madstat} shows the PDFs when the visibilities having amplitude larger than $5\times \rm{MAD}$ are flagged. We find that the large outlier values are flagged by the $5\times \rm{MAD}$ filter. However, some outliers still remain in the data which are flagged by the $3\times \rm{MAD}$ filter (right panel). We have finally used the data after applying the $3\times \rm{MAD}$ filter for which the PDF of the visibility data closely matches with a Gaussian distribution with $\sigma_{N}=0.33\,{\rm Jy}$ that  gives an estimate of noise in the visibility data. The change in the flagged data percentage and the r.m.s. of the visibility data due to different thresholds of the MAD filter are shown in Figure~\ref{fig:flagrms}. Although the fagged  percentage goes up when we apply a $3\times \rm{MAD}$ filter, we have the advantage that the noise level $(\sigma)$ drop by $\sim 62 \%$ .  Further, the visibility statistics is now consistent with a Gaussian distribution, which is expected in our data that is noise dominated.

After applying the $3\times \rm{MAD}$ filter, we have carried out IM analysis on the $100\,\rm{MHz}$ bandwidth data, which span the redshift range $z=1.9-2.6$. The data have $N_c = 4096$ frequency channels where the central channel has the frequency $\nu_c = 444 \,\rm{MHz}$ ($z_c=2.2$). We note that the present analysis is restricted to the dense and nearly uniform baseline range $\mathbf{U} \leq 500 \, \lambda$ (\citetalias{P22}).



\section{The Wideband TGE}
\label{sec:wtge}

The Tapered Gridded Estimator (TGE) which we use here estimates the MAPS \maps{} from the measured visibility data \citep{Bh18, Pal20}. In the present work, we use the `Cross' TGE, which cross-correlates the visibilities measured in the RR and LL polarization states. The formalism of the Cross TGE is detailed in \citetalias{AE23} (also \citetalias{AE23b}), and here we briefly summarize it to highlight the changes which we have incorporated specifically for the wideband analysis.

We have introduced a rectangular grid in $uv$ space, and the first step in TGE is to calculate the convolved-gridded visibility at each grid point $\mathbf{U}_g$ 
\begin{equation}
    \V_{cg}^{x}(\nu_a) = \sum_i  \tilde{w}\left(\mathbf{U}_g - \mathbf{U}_i \right) \, \V{}_i^{x}(\nu_a) \, F_i^{x}(\nu_a) \,.
	\label{eq:vcgx}
\end{equation}
Here $\V{}_i^{x}(\nu_a)$  denotes a visibility measured at a frequency $\nu_a$,  polarization state $x$ and the baseline $\mathbf{U}_i = (\nu_a/c) \times \mathbf{d}_i$ where $\mathbf{d}_i$ refers to  the antenna separation in  length units.  {\it Regarding a fixed antenna pair, the fact that the baseline $\mathbf{U}_i$ scales with frequency $\nu_a$ (baseline migration) is a feature that we have specifically incorporated here for the wideband analysis.}  We note that this was not incorporated in the TGE versions used for the earlier works. The other terms retain the same meaning as in our earlier work; $F_i^{x}(\nu_a)$ is $0$ if the frequency channel is flagged and it is $1$ otherwise, 
and $\tilde{w}(\mathbf{U})$ is the Fourier transform of the tapering window function ${\mathcal W}(\theta)=e^{-\theta^{2}/[f \theta_{0}]^2}$ in which we have used $f=0.6$ for the present work (following \citetalias{P22}).

After accounting for baseline migration, we cross-correlate these convolved-gridded visibilities $\V_{cg}^{RR}(\nu)$ and $\V_{cg}^{LL}(\nu)$ in exactly the same way as done in \citetalias{AE23b} (equation 3). We normalize the estimator using simulated visibilities corresponding to the Gaussian random field with $C_{\ell}(\nu_a,\nu_b)=1$. {\it The simulated visibilities also incorporate the baseline migration and the frequency-dependent variation of the PB pattern of the telescope}. We refer to Section 3.1 of \citetalias{AE23b} for the details.

We use the TGE  to estimate \mapsb{} at each grid point $\mathbf{U}_g = \ellb_g / 2\pi$. Following the notation of \citetalias{AE23b}, here we use the vector $\ellb$ to refer to a grid point.  
The $21$-cm signal is isotropic on the sky plane, and for the signal $C_{\ellb}(\nu_a, \nu_b) = C_\ell(\nu_a, \nu_b)$, where $\ell = \mid \ellb \mid$ corresponds to an angular multipole. 
However, the measured \mapsb{} are expected to be dominated by foregrounds, which are different for different  $\ellb$. We found that it is beneficial to remove the foregrounds from each $\ellb$ independently and impose the isotropy of the signal only after foreground removal. 
Further, the post-EoR $21$-cm signal is assumed to be ergodic (statistically homogeneous) over a significantly large redshift range \citep{Deb16, Not, Zafar, Rhee2018}. This ergodicity along the frequency axis allows us to assume \mapsb{} to be stationary, i.e., \mapsb{}  can be entirely described by the \clb{}, where $\Delta\nu = \mid \nu_a - \nu_b \mid$. We use the estimated \clb{} in the subsequent sections.  

The system noise present in the visibility data contributes to the statistical fluctuations of the measured  \clb{}.  To estimate this, we have simulated multiple $(50)$ realizations of visibilities containing  Gaussian random noise with zero mean  and variance $\sigma_{N}^2$ where $\sigma_{N}$ is the r.m.s. of the actual visibility data. We have applied TGE on each noise-only simulation, and used these to estimate the system noise contribution to  the r.m.s. statistical fluctuations of \clb{}.

\section{Foreground Removal}
\label{sec:FGremoval}

In this section, we first briefly describe the foreground removal techniques introduced in Section~3 of \citetalias{AE23b} and show the results from the present analysis. The idea is based on the distinctly different  behaviour of the $21$-cm signal as compared to the foregrounds. The signal is expected to be predominantly localized within a typical $\Delta \nu $ range of  $ 0 - 0.5 \, \rm{MHz}$, and its amplitude drops substantially and is close to zero at large $\Delta \nu$ \citep{BA5, BS01}. On the other hand, the foregrounds generally vary smoothly and remain correlated over a large $\Delta \nu$ range. Following \citet{ghosh1}, we adopt a two-step procedure to remove the foregrounds from the estimated \clb{}.  First, we identify   a range   $\Delta \nu > \left[ \Delta \nu \right]$ where the  $21$-cm signal is predicted to be  negligible, and model \clb{} in this range  as a combination of foregrounds $\left[C_{\ellb}(\Delta\nu)\right]_{\rm FG}$ and noise
\begin{equation}
    C_{\ellb}(\Delta\nu) = \left[C_{\ellb}(\Delta\nu)\right]_{\rm FG} + [\textrm{Noise}] \,.
    \label{eq:polyfitmodel} 
\end{equation}
The idea is to use the \clb{}  measured in the range $\Delta \nu > \left[ \Delta \nu \right]$ to estimate  the foreground model $\left[C_{\ellb}(\Delta\nu)\right]_{\rm FG}$.
 For the present analysis, we have chosen a fixed value $\left[ \Delta \nu \right] = 0.6 \, \rm{MHz}$  and used the range $0.6 < \Delta \nu < 3.66 \,{\rm MHz}$ to model the foregrounds.  In the second step, we extrapolate the foreground model to predict $ \left[C_{\ellb}(\Delta\nu)\right]_{\rm FG}$ in the range $\Delta \nu \leq \left[ \Delta \nu \right]$ and subtract this out from the measured \clb{}. 
Restricting the subsequent analysis to $\Delta \nu \leq \left[ \Delta \nu \right]$, we expect 
the residual 
\begin{equation}
    \left[C_{\ellb}(\Delta\nu)\right]_{\rm res} = C_{\ellb}(\Delta\nu) - \left[C_{\ellb}(\Delta\nu)\right]_{\rm FG} 
    \label{eq:residual}
\end{equation}
to be a combination of the 21-cm signal and noise
\begin{equation}
     \left[C_{\ellb}(\Delta\nu)\right]_{\rm res}= \left[C_{\ell}(\Delta\nu)\right]_{\rm T} + [\textrm{Noise}]
\end{equation}
and we use this to constrain the  $21$-cm signal $\left[C_{\ell}(\Delta\nu)\right]_{\rm T}$.  

\begin{figure*}
    \centering
    \includegraphics[width=\textwidth]{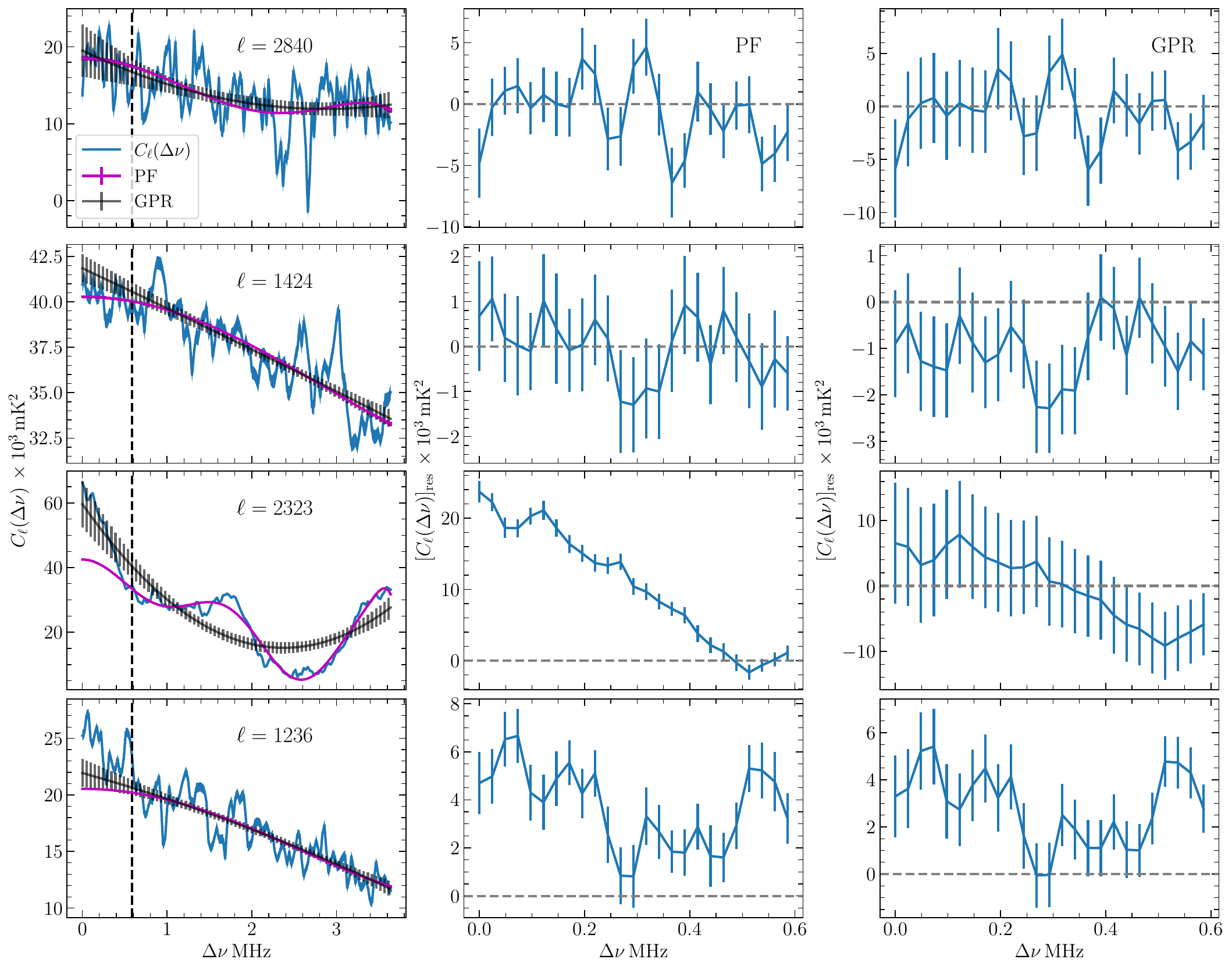}
    \caption{Row-wise, four cases are shown from top to bottom: (1) PF and GPR both work, (2) PF works, but GPR fails, (3) GPR works, but PF fails,  (4) PF and GPR both fail. Column-wise from left to right, the first column shows measured \clb{}, polynomial fit (magenta) and GPR fit (black). The error bars are the $2\sigma$ uncertainties in the GPR foreground model predictions. Note that the errors  for both the measured \clb{} and the PF foreground model predictions are  rather small, and these would not be visible in the panels of the first  column. 
    The vertical lines (black) are drawn at $[\Delta\nu] = 0.6 \, {\rm MHz}$.  The second and third columns show $\left[C_{\ellb}(\Delta\nu)\right]_{\rm res}$ in the range $\Delta \nu \le [\Delta \nu]$ for PF and GPR, respectively. Here, the $2\sigma$ error bars represent the total errors, which combine the errors due to system noise and foreground modelling, and the gray horizontal lines show zero. Note that the error predictions are scaled up by the factor $\sigma_{\rm Est}$ to account for the excess variance in the data which we have discussed later in the paper.}
    \label{fig:cl}
\end{figure*}

Figure \ref{fig:cl} shows \clb{} for a few values of $\ellb$. We see that the $\Delta\nu$ dependence exhibits a smooth, slowly varying pattern superimposed on which we have a combination of rapid and slow oscillations.  
Our aim here is to use the range $\Delta \nu > \left[ \Delta \nu \right]$ to model the smooth, slowly varying $\Delta\nu$ dependence. 
We have also tried modelling the oscillatory components, however, in many cases, we find that the extrapolation 
introduces extraneous features in the range $\Delta \nu \leq \left[ \Delta \nu \right]$. Even in the cases where the extrapolation works, including the oscillatory components leads to a very large signal loss. Appendix~\ref{app:covariance} provides an example of the consequences of including an oscillatory component in the foreground model.

The foreground removal technique described above has been implemented in two different approaches. In our first approach, polynomial fitting (PF), we model the foregrounds using   
\begin{equation}
    \left[C_{\ellb}(\Delta\nu)\right]_{\rm FG} = \sum_{m=0}^n a_{2m} \,\, (\Delta\nu)^{2m} \, .
    \label{eq:FGmodel}
\end{equation}
We use maximum likelihood to estimate the best-fit polynomial coefficients $a_{2m}$ and their error covariance,  and use these to obtain $\left[C_{\ellb}(\Delta\nu)\right]_{\rm FG}$  and the foreground modelling errors in the range $\Delta \nu \leq \left[ \Delta \nu \right]$. Note that we have restricted the value of $n$ to the range $0\leq n \leq8$ to model only the smooth, slowly varying component from the measured  \clb{}.

In our second approach we have used Gaussian Process Regression (GPR; \citealt{RW}), which is non-parametric, to model and predict the foreground. This was first implemented in \citetalias{AE23b}, where the reader can find more details. Note that we have used the \textsc{george} \citep{george} library of \textsc{python} for the entire GPR analysis presented here. In GPR, we model $\left[C_{\ellb}(\Delta\nu)\right]_{\rm FG}$ as a Gaussian Process (GP)
\begin{equation}
    \left[C_{\ellb}(\Delta\nu)\right]_{\rm FG} \sim \mathcal{GP} \left[ 0,\, k_{\rm{FG}}(\Delta\nu_m, \Delta\nu_n) \right] \,,
    \label{eq:FGmodel_GP}
\end{equation}
with a zero mean  and  covariance function (kernel)  $ k_{\rm{FG}}(\Delta\nu_m, \Delta\nu_n)$. The idea is to use the \clb{}  measured in the range $\Delta \nu > \left[ \Delta \nu \right]$ to estimate the kernel, which models the smooth, slowly varying foreground component $ \left[C_{\ellb}(\Delta\nu)\right]_{\rm FG}$. We subsequently use this kernel to predict $ \left[C_{\ellb}(\Delta\nu)\right]_{\rm FG}$ in the range $\Delta \nu \leq  \left[ \Delta \nu \right]$. We have tried  different possible forms for the kernel, and found that the polynomial  kernel 
\begin{equation}
    k_{\rm{FG}}(\Delta\nu_m, \Delta\nu_n) = c_1 \, (\Delta\nu_m \cdot \Delta\nu_n + b )^{P}  
    \label{eq:kernel}
\end{equation}
is well  suited for our analysis. Here, the constants $c_1$ and $b$ are hyper-parameters  whose  optimal value  have been determined  by maximizing the log marginal likelihood of the GP, and  $P$, which is not a hyper-parameter, denotes the order of the polynomial kernel.  The polynomial kernel is found to model the overall smooth $\Delta \nu$ dependence expected of the foregrounds and causes a minimal signal loss. We have considered different values of $P$ and found that larger values provide a better fit to the foregrounds in the range $\Delta \nu > [\Delta \nu]$ but at the cost of a larger signal loss. Considering this, we have used $P = 2$  for the entire analysis. We have presented the results from a selected choice of kernels in Appendix~\ref{app:covariance}. 

The GPR predicts the foregrounds $\left[C_{\ellb}(\Delta\nu)\right]_{\rm FG}$ and also the  foreground modelling errors  in the range $\Delta \nu \leq [\Delta \nu]$. For both PF and GPR, we have added the variance of foreground modelling errors to the estimated system noise variance to estimate the total error variance for $\left[C_{\ellb}(\Delta\nu)\right]_{\rm res}$.

We find that foreground removal does not work for all the $\ellb$. In order to identify the $\ellb$ where the foreground removal fails, we have quantified the amplitude of the residual $ [C_{\ellb}(\Delta\nu)]_{\rm{res}}$ using a single parameter {\bf A} whose definition we present below. We have fitted $ [C_{\ellb}(\Delta\nu)]_{\rm{res}}$ using the predicted 21-cm signal $ \left[C_{\ellb}(\Delta\nu)\right]_T$ (equation~5 of \citetalias{AE23b}; also see \citealt{BA5})
\begin{align}
    \left[C_{\ellb}(\Delta\nu)\right]_T =  \left[\Omega_{\HI} b_{\HI}\right]^{2}  & \frac{\bar{T}^{2}}{\mathrm{\pi} r^2} \int_{0}^{\infty}   d k_{\parallel} \cos(k_{\parallel}r^{\prime}\Delta\nu)  \, \nonumber \\ 
    \times & \, \mathrm{sinc}^2(k_{\parallel}r^{\prime}\Delta\nu_c/2) \,     
    P_m(\mathbf{k_{\perp}}, k_{\parallel}) \, 
    \label{eq:cl_Pk_sinc_p4}
\end{align}
where $P_{m}(\mathbf{k_{\perp}}, k_{\parallel})$ is the dark matter power spectrum at the wave vector  $\mathbf{k}$ whose line-of-sight and perpendicular components are $k_{\parallel}$ and $\mathbf{k_{\perp}} = {\ellb}/{r}$ respectively, and the $\rm{sinc}$ function accounts for the finite channel width $\Delta\nu_c$. 
The comoving distance $r$, its derivative with respective to frequency $r^\prime$ and the 21-cm mean brightness temperature $\bar{T}$ have the values 5598~Mpc, $9.70 \,{\rm Mpc \, MHz^{-1}}$ and $400 \, {\rm mK}$, respectively, at $z_c=2.2$. 
We define the parameter ${\bf A} \equiv \left[\Omega_{\HI} b_{\HI}\right]^{2}$ which quantifies the overall amplitude of the predicted 21-cm signal $ \left[C_{\ellb}(\Delta\nu)\right]_T$. 
Here we treat ${\bf A}$ as a free parameter and use it to effectively quantify the amplitude of  $ [C_{\ellb}(\Delta\nu)]_{\rm{res}}$, with ${\bf dA}$ denoting the predicted uncertainties. We have initially tried the flagging criteria $\mid {\bf A} \mid/{\bf dA} > 3$ to identify and remove the $\ellb$ values where foreground subtraction fails. The aim here is to get those $\ellb$ values where the residuals are largely consistent with noise. We use all unflagged $\ellb$ values to improve the SNR.  In our analysis, we have also considered several different flagging criteria such as $\mid {\bf A}\mid / {\bf dA} > 2$, $4$, and $5$. 
We find that a few more $\ellb$ values are flagged if we choose $\mid {\bf A}\mid / {\bf dA} > 2$, whereas we pick up a few $\ellb$ values having residual foregrounds if we choose the flagging criteria $\mid {\bf A}\mid / {\bf dA} > 5$. The results from all the different flagging criteria considered here are presented in Appendix~\ref{app:flag}. We obtain the best results if we use the flagging criteria $\mid {\bf A} \mid/{\bf dA} > 4$ for PF and $\mid {\bf A} \mid/{\bf dA} > 3$ for GPR. We have adopted these values for the final results presented in all the subsequent analyses.


We have separately applied PF and GPR on all the $\ellb$. We find  $43$ and $47$ unflagged $\ellb$  for  PF and GPR, respectively. However, these are largely mutually exclusive,  and only  $6$  of the $\ellb$  are common to both PF and GPR. Considering four representative cases, the four rows of  Figure~\ref{fig:cl} show the results for foreground removal. The first row (from the top) considers a case where both PF  and GPR successfully model and remove the foregrounds. The left column shows the  measured \clb{},  along with $\left[C_{\ellb}(\Delta\nu)\right]_{\rm FG} $ for  both PF  and GPR. The vertical dashed line shows $[\Delta \nu]$, and the middle and right columns,  which show $\left[C_{\ellb}(\Delta\nu)\right]_{\rm res}$  for PF and GPR respectively,  are restricted to the range $\Delta \nu \le [\Delta \nu]$. 
For both PF and GPR, the residuals are found to be scattered around zero. 
Note that the $2 \sigma$ error bars shown for  $\left[C_{\ellb}(\Delta\nu)\right]_{\rm res}$ combine the contributions from system noise and the uncertainties in the foreground model predictions. The second row considers a case where PF works but GPR fails because it over predicts $\left[C_{\ellb}(\Delta\nu)\right]_{\rm FG} $ whereby   $\left[C_{\ellb}(\Delta\nu)\right]_{\rm res}$  
is consistently below zero (right panel).   The third row shows a case where GPR works but PF fails because it under predicts $\left[C_{\ellb}(\Delta\nu)\right]_{\rm FG} $ and 
$\left[C_{\ellb}(\Delta\nu)\right]_{\rm res}$   is consistently above  zero (middle panel). The last row shows a case where both PF and GPR fail because they  under predict $\left[C_{\ellb}(\Delta\nu)\right]_{\rm FG} $, and the values of $\left[C_{\ellb}(\Delta\nu)\right]_{\rm FG} $ are consistently above zero. We draw the reader's attention to the oscillatory features which are visible   in the range $\Delta \nu > [\Delta \nu]$ for all the panels in the first column. This is particularly prominent in the third row. It is also quite obvious that our foreground model predictions do not match these oscillatory features. As demonstrated in  Appendix~\ref{app:covariance}, including an oscillatory component in the foreground model increases the 21-cm signal loss, and in some cases it may also worsen the extrapolation in the range $\Delta \nu \leq [\Delta \nu]$. Based on this, we have only modelled the smooth,  slowly varying $\Delta \nu$ dependence and avoided modelling the oscillations.

For the subsequent analysis, we have discarded the  $\ellb$  grids  where both PF and GPR fail to remove the foregrounds. We have separately analysed three sets, namely PF, GPR and Combined, which respectively contain $43$,   $47$ and $84$ $\ellb$ grids.   For the $6$ grids where both PF and GPR work, we have used GPR, which was found to yield tighter constraints.  The $\mid \ellb \mid$  values roughly span the range $900 \lesssim \mid \ellb \mid \lesssim 3100$.  

Foreground removal is expected to also remove some of the  21-cm signal.  Following \citetalias{AE23b} where we have taken a very conservative approach, we have applied the foreground removal technique on $\left[C_{\ell}(\Delta\nu)\right]_{\rm T}$ the expected 21-cm signal and used this to estimate the loss in the 21-cm signal.  Considering the power spectrum, we find that the signal loss is most prominent  at low $k$ where it has value $\sim 40-50 \%$ at  $k \sim 0.2 \, {\rm Mpc}^{-1}$ and 
$\sim 20 \%$ at  $k \sim 0.4 \, {\rm Mpc}^{-1}$. The signal loss is less than $10\%$ at larger $k$. We have corrected for this estimated signal loss in all the results presented in the subsequent sections.

\section{The Cylindrical PS}
\label{sec:cylindricalps}

The cylindrical PS $P(k_{\perp}, k_{\parallel})$  is related to MAPS \cl{} via a Fourier transform along $\Delta\nu$ \citep{KD07}
\begin{equation}
    P(\kb, k_{\parallel})= r^2\,r^{\prime} \int_{-\infty}^{\infty}  d (\Delta \nu) \, \mathrm{e}^{-i  k_{\parallel} r^{\prime} \Delta  \nu}\, C_{\ellb}(\Delta \nu) \,. 
\label{eq:cl_Pk}
\end{equation}
We have used a maximum likelihood estimator (MLE; \citealt{Pal20}) to estimate the cylindrical PS  $P(\kb, k_{\parallel})$ from \clb{}. 
\begin{figure*}
    \centering
    \includegraphics[width=\textwidth]{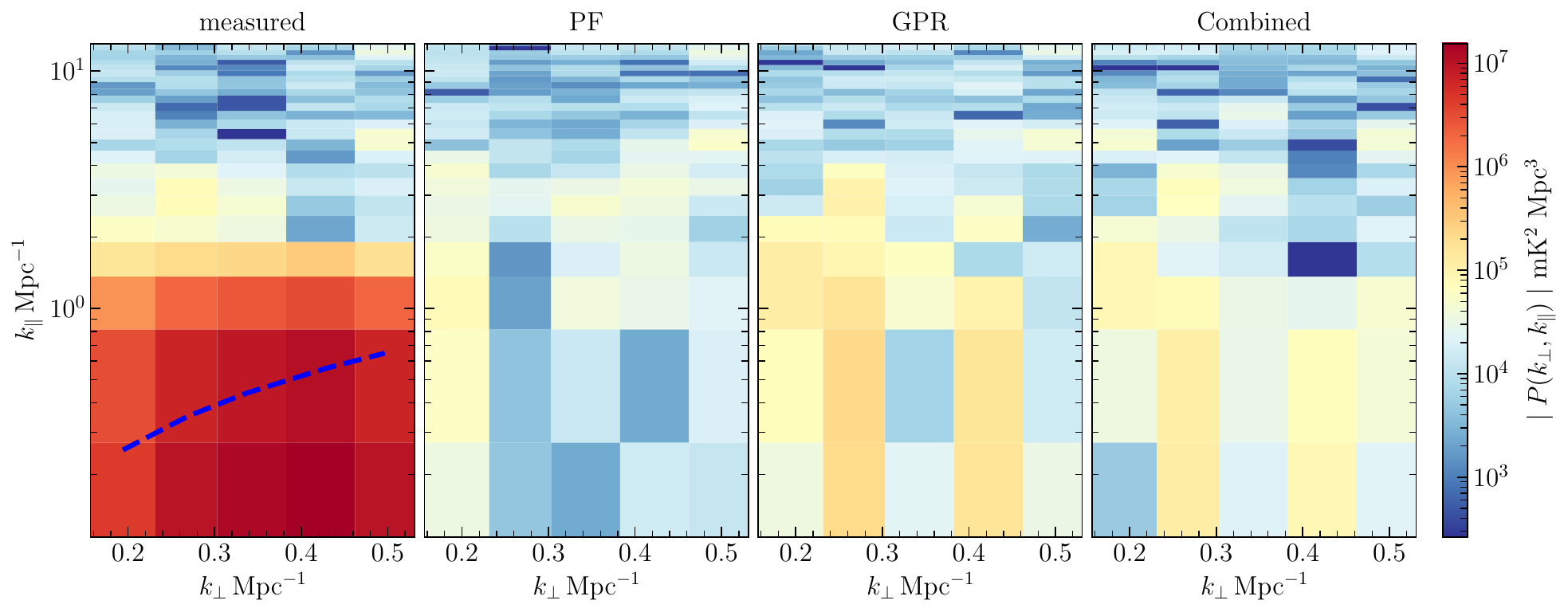}
    \caption{The left panel shows $\mid P(k_{\perp}, k_{\parallel}) \mid$ before foreground removal. The second, third and fourth panels show $\mid P(k_{\perp}, k_{\parallel}) \mid$ after foreground removal for PF, GPR and Combined, respectively. The blue dashed line shows the predicted foreground wedge boundary.}
    \label{fig:cylps}
\end{figure*}

\begin{figure}
    \centering
    \includegraphics[width=\columnwidth]{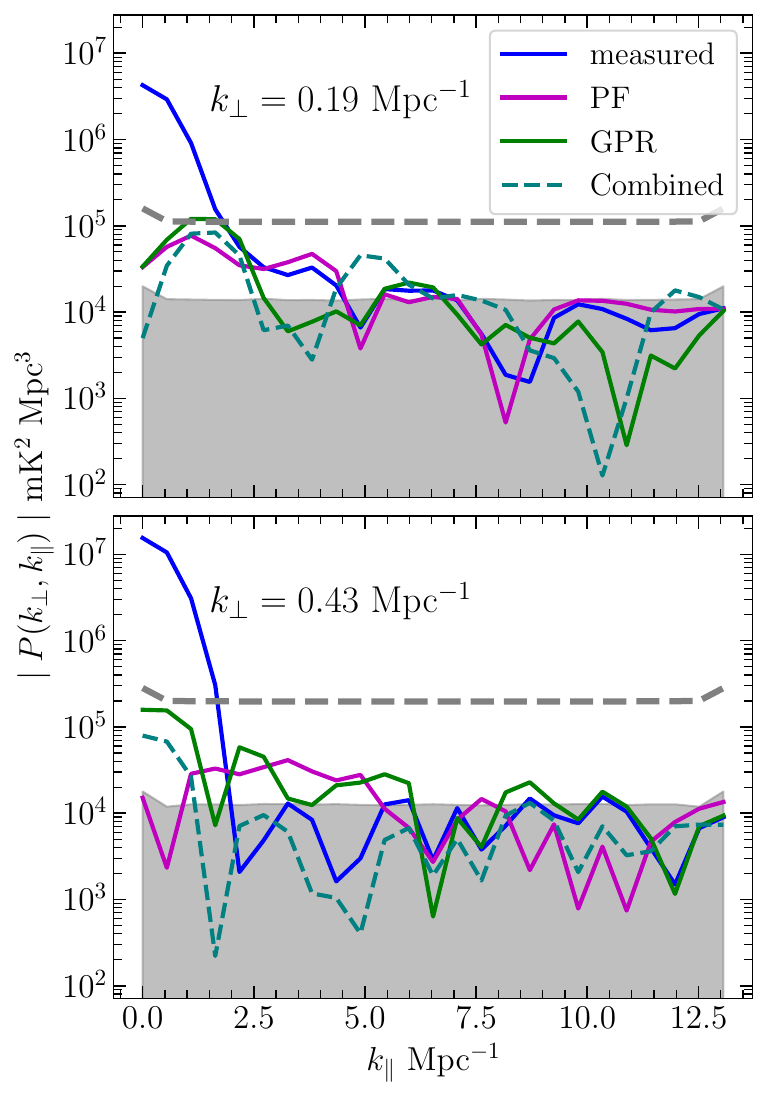}
    \caption{This shows the cylindrical PS $\mid P(k_{\perp}, k_{\parallel}) \mid$ for  fixed $k_\perp$  as a function of $k_\parallel$. The blue curves show $\mid P(k_{\perp}, k_{\parallel}) \mid$ before foreground removal. The magenta, green  and teal curves show $\mid \left[ P(k_{\perp}, k_{\parallel})\right]_{\rm res} \mid$ obtained from PF, GPR and Combined, respectively. The shaded regions show the estimated $2\sigma$ statistical fluctuations  from system noise only. The grey dashed horizontal curve shows the total $2\sigma$ errors (system noise + foreground modelling) for Combined.}
    \label{fig:pkslice}
\end{figure}

Here, for the purpose of visualising the results (only Figures~\ref{fig:cylps} and \ref{fig:pkslice}), we have binned the $C_{\ellb}(\Delta\nu)$ values into 5 equally spaced annular bins in the $uv$-plane,  and we show the binned cylindrical PS \pk{}. The left panel of Figure~\ref{fig:cylps} shows $\mid P(k_{\perp}, k_{\parallel}) \mid$ before foreground removal.   Considering the full $(k_{\perp}, k_{\parallel})$ plane, the values of $\mid P(k_{\perp}, k_{\parallel})\mid$ are found to be in the range $\sim 10^3 - 10^7  {\rm mK^2}$. We find that most of the large values $\mid P(k_{\perp}, k_{\parallel}) \mid > 10^5  {\rm mK^2}$ are restricted to the region $k_{\parallel} \leq 2 \, {\rm Mpc^{-1} }$  which extend slightly beyond the theoretically predicted foreground wedge boundary (blue dashed line).  This region is entirely foreground-dominated. Considering the  region  $k_{\parallel} \geq 2 \, {\rm Mpc^{-1} }$,  we find $\mid P(k_{\perp}, k_{\parallel}) \mid$ has values $\sim 10^3 - 10^5  {\rm mK^2}$. It is reasonable to expect that the power in this higher $k_{\parallel}$ region is largely due to noise.

The second panel of Figure~\ref{fig:cylps}  shows $\mid P(k_{\perp}, k_{\parallel}) \mid$ corresponding to $\left[C_{\ell}(\Delta\nu)\right]_{\rm res}$  obtained from foreground removal using  PF. Here $\mid P(k_{\perp}, k_{\parallel}) \mid$ has values $\sim 10^3 - 10^5  {\rm mK^2}$ throughout the $(k_{\perp}, k_{\parallel})$ plane. We find that the power in the $k_{\parallel} \leq 2 \, {\rm Mpc^{-1} }$ region has reduced by nearly $2$ orders of magnitudes after foreground removal. The third and fourth panels show $\mid P(k_{\perp}, k_{\parallel}) \mid$ for GPR and Combined respectively.  Here also the results are similar to those for PF, and the entire  $(k_{\perp}, k_{\parallel})$ plane appears to be free of foreground contamination. 
Comparing the PS for PF and GPR, we find that GPR  results in  slightly higher values of $\mid P(k_{\perp}, k_{\parallel}) \mid$.  It may however be noted that GPR also has larger  uncertainties in the predicted foreground model $ \left[C_{\ellb}(\Delta\nu)\right]_{\rm FG}$, and the total $2 \sigma$ error bars also are larger compared to PF  (Figure \ref{fig:cl}). It may be possible to reduce these by varying the value of $P$ (polynomial order of the covariance)  which has now been held fixed at $P=2$ for GPR. However, this could enhance the 21-cm signal loss, and we have not tried this here.  The $\mid P(k_{\perp}, k_{\parallel}) \mid$ for the Combined set is also very similar to that of PF and GPR. In general, we may conclude that after foreground removal the estimated $P(k_{\perp}, k_{\parallel})$  are largely foreground-free, although some low level residual foregrounds comparable to the noise level may possibly  still be present in the data.

For a closer inspection, we have also considered the variation of $\mid P(k_{\perp}, k_{\parallel}) \mid$ as a function of $k_{\parallel}$  for the individual  $k_\perp$ bins.  Figure~\ref{fig:pkslice} shows $\mid P(k_{\perp}, k_{\parallel}) \mid$ for two representative $k_\perp$ bins, $k_\perp = 0.19\,{\rm Mpc^{-1}}$ and $0.43\,{\rm Mpc^{-1}}$. The grey-shaded region shows the estimated $2\sigma$ noise level from system noise only.  The blue curves show $\mid P(k_{\perp}, k_{\parallel}) \mid$ corresponding to the measured \cl{}.  The values of $\mid P(k_{\perp}, k_{\parallel})\mid$ are found to be in the range $\sim 10^3 - 10^7  {\rm mK^2}$. The large values $\mid P(k_{\perp}, k_{\parallel}) \mid > 10^5  {\rm mK^2}$ are mostly confined in  the region  $k_{\parallel} \leq 2 \, {\rm Mpc^{-1} }$. In the region $k_{\parallel} > 2 \, {\rm Mpc^{-1} }$, the values of $\mid P(k_{\perp}, k_{\parallel}) \mid$ are either slightly above the $2 \sigma$ system noise level or consistent with it. 

The magenta, green and teal curves show $\mid P(k_{\perp}, k_{\parallel}) \mid$ after foreground removal for PF, GPR and Combined, respectively. The values of $\mid P(k_{\perp}, k_{\parallel}) \mid$ are in the range $\sim 10^3 - 10^5  {\rm mK^2}$. We find that the dominant foreground contamination having large values $(\mid P(k_{\perp}, k_{\parallel}) \mid > 10^5  {\rm mK^2})$ are removed by PF and GPR.  Although there is a substantial reduction, we  find that the values of $\mid P(k_{\perp}, k_{\parallel}) \mid$ are still above the  $ 2 \sigma$  system noise level, particularly at small $k_{\parallel}$. However, it is necessary to note that in addition to the system noise,  the  error estimates after foreground subtraction also have a contribution from the uncertainties in the foreground model. Further, the latter contribution is expected to differ depending on whether we consider PF, GPR or Combined. The dashed horizontal curve in Figure~\ref{fig:pkslice} shows the total $2 \sigma$ errors for Combined.  We find that all the $\mid P(k_{\perp}, k_{\parallel}) \mid$  values are below this curve after foreground subtraction.

We now examine the statistics of the estimated  $P(\kb, k_{\parallel})$ through the quantity $X$  (\citealt{Pal20}) 
\begin{equation}
    X=\frac{P(\kb, k_{\parallel})}{\delta P_{N}(\kb,\,k_{\parallel})}
    \label{eq:xstat}
\end{equation}
where $\delta P_{N}(\kb,\,k_{\parallel})$ is the total error due to system noise and the foreground modelling. $X$ is expected to have a symmetric distribution around a mean $\mu = 0$ with a standard deviation $\sigma_{\rm Est} = 1$ if the estimated values of $P(\kb, k_{\parallel})$ are consistent with  the predicted uncertainties $\delta P_{N}(\kb,\,k_{\parallel})$.  Any residual foregrounds are expected to skew the distribution of $X$ towards a positive value. On the other hand, negative systematics which could potentially bias the estimated $P(\kb, k_{\parallel})$ towards a low value would skew the distribution of $X$ towards a negative  value. However,  if the distribution of $X$ is symmetric with  $\sigma_{\rm Est} > 1$, we can say that the actual uncertainty in the estimated  $P(\kb, k_{\parallel})$ are larger than the predicted errors $\delta P_{N}(\kb,\,k_{\parallel})$.

\begin{figure*}
    \centering
    \includegraphics[width=\textwidth]{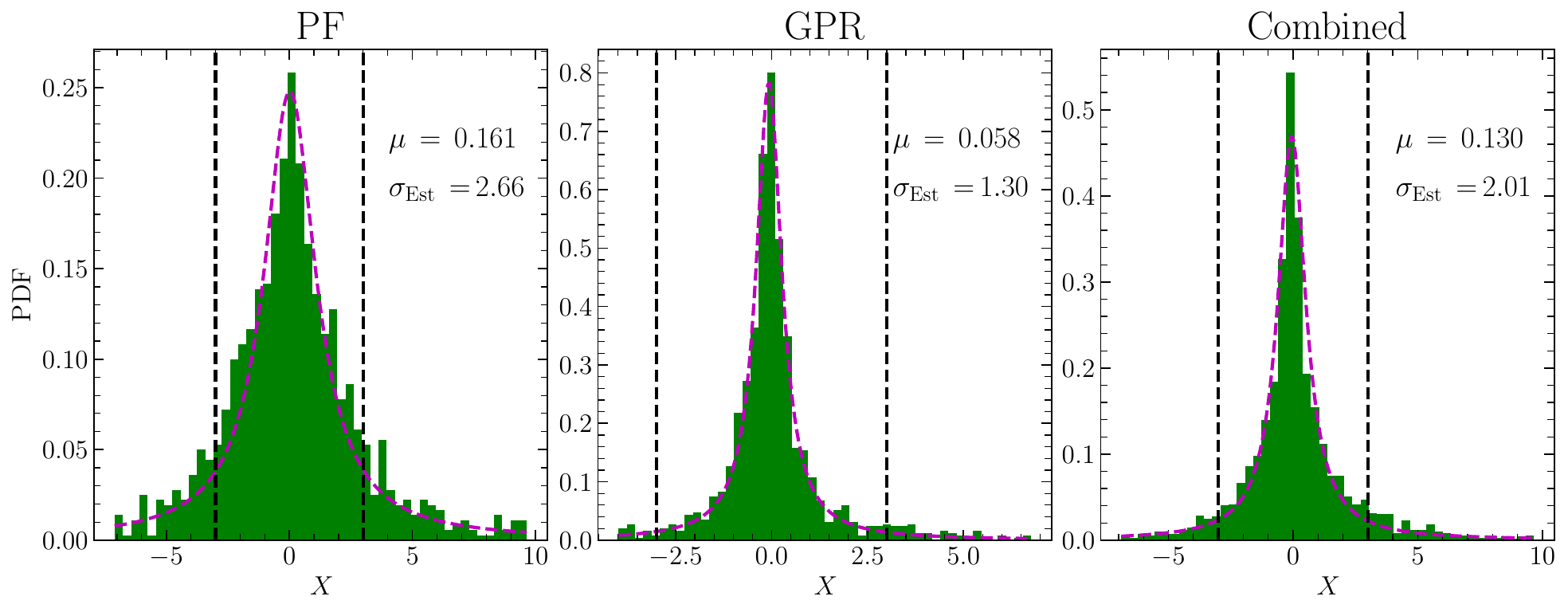}
    \caption{The probability density functions (PDF) of $X$ for PF, GPR and Combined. Lorentzian fits of the PDFs are shown with the magenta lines.  The black vertical lines show $\mid X \mid \leq 3$ within which $80-90\%$ values are found to reside. }
    \label{fig:noisestat}
\end{figure*}

Figure~\ref{fig:noisestat} shows the probability density function (PDF) of $X$ for PF (left), GPR (middle) and Combined (right). Considering PF, we see that the PDF is largely symmetric around $0$.  The majority $(\sim 80-90\%)$ of the values of $X$  lie in the central part of the distribution $ \mid X \mid \le 3$ which is demarcated by the black dashed vertical lines. We find $\sim2-3\%$ of the samples to have slightly larger values $(\mid X \mid >5)$, which are possibly noise or trace amounts of unsubtracted foregrounds or systematics.
The PDF of $X$ for GPR looks similar to PF, but the values of $X$ are found to be somewhat smaller. These smaller values are due to the fact that the error predictions from GPR are larger than those of PF. The  PDF of $X$ for Combined also shows a similar trend as that of PF.

We find that the mean $(\mu)$ of $X$ is $\sim0$ for all three cases, which suggests that the foregrounds have largely been removed. Considering $\sigma_{\rm Est}$, we find this to have values  2.66, 1.30 and 2.01 for PF, GPR and Combined respectively. The difference in $\sigma_{\rm Est}$ between  PF and GPR is due to the different error predictions from these two methods. The value $\sigma_{\rm Est} > 1$ indicates that we have some excess variance as compared to our error predictions $\delta P_{N}(\kb,\,k_{\parallel})$. We note that an excess variance after foreground removal has previously been noted in other low-frequency observations as well \citep{mertens20, Pal20}. The excess variance possibly arises because of imperfect calibration, low-level RFIs, inaccurate point source subtraction, etc. \citep{Gan2022}. Note that we have found  a larger value $\sigma_{\rm Est} = 4.77$ in our earlier analysis (\citetalias{AE23}, \citetalias{AE23b}). The reduction in the excess variance in the present analysis is possibly due to the MAD filter we have used to flag the outliers in the visibility data (Section~\ref{sec:data}). It is further possible that the fourfold increase in the bandwidth has also contributed to this lower excess variance. Although the exact reason for the improvement is not known, it is expected that the reduction in the excess variance will tighten the constraints on the $21$-cm PS. Following \citetalias{AE23b}, for all the subsequent analyses, the error predictions are scaled up by the factor $\sigma_{\rm Est}$ to account for the excess variance in the data.

Similar to the earlier works, here also we find that the PDFs of $X$ closely follow a Lorentzian distribution. The best-fit Lorentzian PDFs, which are shown with the magenta lines (Figure~\ref{fig:noisestat}), are found to have a peak location $x_0 \sim 0$ and spread $\gamma = $ 1.28, 0.40 and 0.68 for PF, GPR and Combined respectively.

\section{The Spherical PS}
\label{sec:BMLE}

We have estimated the spherical PS $P(k)$ directly from $[C_{\ell}(\Delta\nu)]_{\rm res}$ using the MLE developed in \citetalias{AE23} (also see \citetalias{AE23b}). The values of $P(k)$ are estimated at  $7$ spherical bins which span the range  $0.2 < k < 6.4 \, \rm{Mpc}^{-1}$. We used the estimated $P(k)$ values to obtain the mean squared brightness temperature fluctuations $\Delta^2(k)\equiv {k^{3}}P(k)/{2\pi^2}$. 

\begin{figure*}
    \centering
    \includegraphics[width=\textwidth]{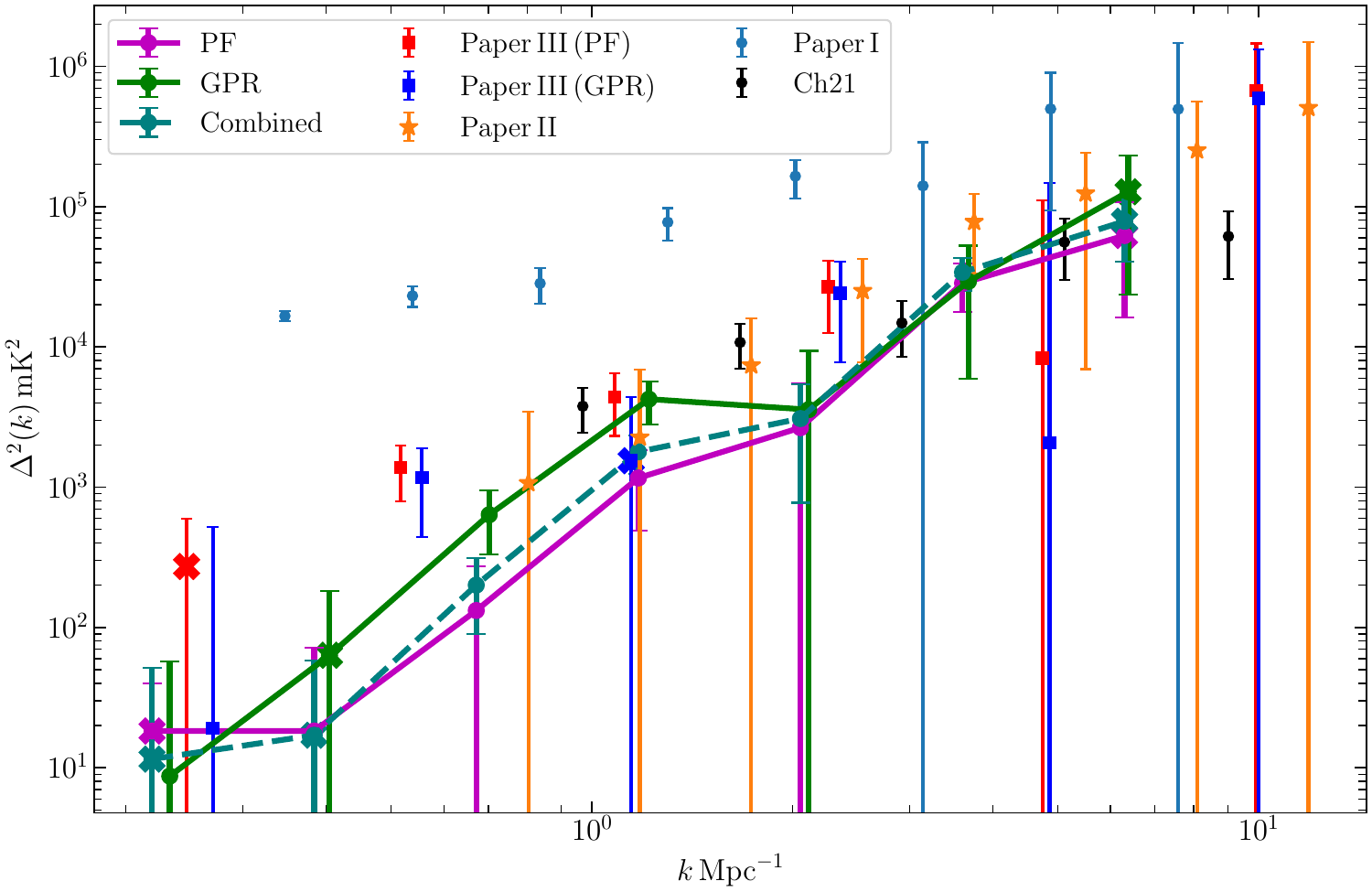}
    \caption{This shows the  measured $\Delta^2(k)$ and  $2 \sigma$ uncertainties where the   magenta, green and teal lines correspond to  PF, GPR and Combined, respectively. The results from
    \citetalias{AE23b} (squares: red - PF and blue - GPR),     \citetalias{AE23} (orange asterisks), \citetalias{P22} (light blue circles) and \citetalias{Ch21} (black circles) are shown for reference. }
    \label{fig:pssph}
\end{figure*}

Figure~\ref{fig:pssph} shows $\Delta^2(k)$ with the $2\sigma$ uncertainties as error bars, which are obtained in different works, including the present one,  using this Band $3$ observation. We first focus on the present work for which the magenta, green and teal lines correspond to the $\Delta^2(k)$ obtained from PF, GPR and Combined respectively. Here, the negative values of $\Delta^2(k)$ are marked with crosses. 

Considering PF, for which the results are also tabulated in Table~\ref{tab:ul_MLE}, we find the values of $\mid \Delta^2(k) \mid$ are roughly in the range $(5)^2$ -- $(250)^2 \, \rm{mK}^2$ for the $k$-range probed here. Here, the $\Delta^2(k)$  values are consistent with noise at $0\pm 2\sigma$ for most of the $k$ bins barring a few which we discuss below.  The $\Delta^2(k)$ values are consistent with noise at $0\pm 3\sigma$ in the fourth and seventh $k$-bins, and it slightly exceeds $5\sigma$ in the sixth bin, possibly due to some surviving foreground contamination.   Considering the $k$ bins where  $\Delta^2(k)$ is negative, we find that the values are within  $0\pm 2\sigma$ and  $0\pm 3\sigma$ for the first and last $k$ bins respectively. This indicates that the PS does not have negative systematics, and our estimates are robust.  We obtain the tightest $2\sigma$ upper limit $\Delta_{\rm UL}^2(k)$ from the first $k$ bin where we have  $\Delta_{\rm UL}^2(k) = (4.68)^2 \, \rm{mK}^2$ at $k = 0.219 \, \rm{Mpc}^{-1}$.  We next consider the results for  GPR (Table~\ref{tab:ul_GPR}) where  we find that  $\mid \Delta^2(k) \mid$ have values in the range $(3)^2 \, \rm{mK}^2$ to $(350)^2 \, \rm{mK}^2$. These values are slightly larger than those we have obtained for PF.  Here, the $\Delta^2(k)$  values are consistent with noise at $0\pm 3\sigma$ for all the $k$ bins except the third and the fourth bins where the SNR is nearly 5. We conclude that some foregrounds remain in the data due to our conservative choice of a low-order polynomial for the covariance function (equation~\ref{eq:FGmodel_GP}). The tightest upper limit from GPR is found to be $\Delta_{\rm UL}^2(k) = (7.56)^2 \, \rm{mK}^2$ at $k = 0.233 \, \rm{Mpc}^{-1}$. For Combined (Table~\ref{tab:ul_comb}), the $\mid \Delta^2(k) \mid$ values are nearly in between those found from PF and GPR, however, the $2\sigma$ error bars are tightened in most of the $k$-bins.  Considering PF, GPR and Combined together, we find that the tightest  upper limit on $\Delta^2(k)$ 
comes from the smallest $k$ bin for PF. We note  that when $\Delta^2(k)$ 
is negative,   we have conservatively set $\Delta^2(k) = 0$  and reported  $2\sigma$  as the upper limit. 

\begin{table}
    \centering
    \caption{Considering PF, this presents the measured $\Delta^2(k)$, corresponding errors $\sigma(k)$, ${\rm SNR} = \Delta^2(k)/\sigma(k)$, and the $2\,\sigma$ upper limits   $\Delta_{\rm UL}^{2}(k)$ and  $[\Omega_{\HI}b_{\HI}]_{\rm UL}$.  }
        \begin{tabular}{cccccc}
        \hline
        \hline
        $k$ & $\Delta^2(k)$ & $1\sigma(k)$ & SNR & $\Delta_{\rm UL}^{2}(k)$ & $[\Omega_{\HI}b_{\HI}]_{\rm UL}$ \\
        $\rm{Mpc}^{-1}$ & $\rm{mK}^2$ & $\rm{mK}^2$ & & $\rm{mK}^2$  & $\times 10^{-2}$\\
        \hline
        $0.219$ &  $-(4.27)^2$ & $(3.31)^2$ & $-1.67$ & $(4.68)^2$ & $1.01$ \\
        $0.384$ &  $(4.28)^2$ & $(5.18)^2$ & $0.68$ & $(8.48)^2$ & $1.41$ \\
        $0.671$ &  $(11.50)^2$ & $(8.34)^2$ & $1.90$ & $(16.47)^2$ & $2.19$ \\
        $1.175$ &  $(34.12)^2$ & $(18.31)^2$ & $3.47$ & $(42.84)^2$ & $4.68$ \\
        $2.057$ &  $(51.56)^2$ & $(37.95)^2$ & $1.85$ & $(74.42)^2$ & $6.88$ \\
        $3.601$ &  $(168.81)^2$ & $(73.63)^2$ & $5.26$ & $(198.35)^2$ & $15.83$ \\
        $6.303$ &  $-(249.54)^2$ & $(151.76)^2$ & $-2.70$ & $(214.62)^2$ & $15.06$ \\
        \hline
    \end{tabular}
    \label{tab:ul_MLE}
\end{table}
\begin{table}
    \centering
    \caption{Same as Table~\ref{tab:ul_MLE},  but for GPR.}
        \begin{tabular}{cccccc}
        \hline
        \hline
        $k$ & $\Delta^2(k)$ & $1\sigma(k)$ & SNR & $\Delta_{\rm UL}^{2}(k)$ & $[\Omega_{\HI}b_{\HI}]_{\rm UL}$ \\
        $\rm{Mpc}^{-1}$ & $\rm{mK}^2$ & $\rm{mK}^2$ & & $\rm{mK}^2$  & $\times 10^{-2}$\\
        \hline
        $0.233$ &  $(2.95)^2$ & $(4.92)^2$ & $0.36$ & $(7.56)^2$ & $1.60$ \\
        $0.404$ &  $-(7.96)^2$ & $(7.71)^2$ & $-1.07$ & $(10.90)^2$ & $1.77$ \\
        $0.702$ &  $(25.23)^2$ & $(12.39)^2$ & $4.15$ & $(30.72)^2$ & $4.01$ \\
        $1.219$ &  $(65.18)^2$ & $(26.87)^2$ & $5.88$ & $(75.45)^2$ & $8.15$ \\
        $2.116$ &  $(59.79)^2$ & $(53.88)^2$ & $1.23$ & $(96.86)^2$ & $8.88$ \\
        $3.675$ &  $(171.36)^2$ & $(108.16)^2$ & $2.51$ & $(229.70)^2$ & $18.25$ \\
        $6.380$ &  $-(357.48)^2$ & $(228.08)^2$ & $-2.46$ & $(322.55)^2$ & $22.57$ \\
        \hline
    \end{tabular}
    \label{tab:ul_GPR}
\end{table}
\begin{table}
    \centering
    \caption{Same as Table~\ref{tab:ul_MLE}, but for Combined.}
        \begin{tabular}{cccccc}
        \hline
        \hline
        $k$ & $\Delta^2(k)$ & $1\sigma(k)$ & SNR & $\Delta_{\rm UL}^{2}(k)$ & $[\Omega_{\HI}b_{\HI}]_{\rm UL}$ \\
        $\rm{Mpc}^{-1}$ & $\rm{mK}^2$ & $\rm{mK}^2$ & & $\rm{mK}^2$  & $\times 10^{-2}$\\
        \hline
        $0.219$ &  $-(3.40)^2$ & $(4.47)^2$ & $-0.58$ & $(6.32)^2$ & $1.37$ \\
        $0.383$ &  $(4.13)^2$ & $(4.51)^2$ & $-0.84$ & $(6.38)^2$ & $1.06$ \\
        $0.671$ &  $(14.16)^2$ & $(7.44)^2$ & $3.63$ & $(17.64)^2$ & $2.34$ \\
        $1.175$ &  $(42.29)^2$ & $(16.51)^2$ & $6.56$ & $(48.31)^2$ & $5.28$ \\
        $2.057$ &  $(55.58)^2$ & $(34.01)^2$ & $2.67$ & $(73.50)^2$ & $6.79$ \\
        $3.601$ &  $(184.67)^2$ & $(66.44)^2$ & $7.73$ & $(207.19)^2$ & $16.54$ \\
        $6.303$ &  $-(280.75)^2$ & $(138.39)^2$ & $-4.12$ & $(195.72)^2$ & $13.73$ \\
    \hline
    \end{tabular}
    \label{tab:ul_comb}
\end{table}

We have also used these upper limits $\Delta_{\rm UL}^2(k)$ to obtain corresponding upper limits on the cosmological \HI{} abundance $[\Omega_{\HI}b_{\HI}]$ at  various  $k$ bins.  The assumption here is that the $21$-cm PS  $P_{T}(\mathbf{k})$  traces the redshift-space matter PS $P^s_{m}(\mathbf{k})$ through 
\begin{equation}
    P_{T}(\mathbf{k}) = \left[\Omega_{\HI} b_{\HI}\right]^{2} \bar{T}^{2} P^s_{m}(\mathbf{k}) \,.
    \label{eq:pT}
\end{equation}
where $\bar{T}$,  the 21-cm mean brightness temperature,  is found  to have a value  $400 \, {\rm mK}$ (equation 6 of \citetalias{AE23b}).
Here, we have obtained  $P^s_{m}(\mathbf{k})$  using \cite{Eisenstein_1998} and ignored the effect of redshift space distortions.   As for the tightest constraints, we find $[\Omega_{\HI}b_{\HI}]_{\rm UL} =  1.01 \times 10^{-2}$  at $k = 0.219 \, \rm{Mpc}^{-1}$ from PF, whereas, we find $[\Omega_{\HI}b_{\HI}]_{\rm UL} =  1.60 \times 10^{-2}$  at $k = 0.233 \, \rm{Mpc}^{-1}$ from GPR. The upper limits obtained from these two independent methods are very close, which gives a nice cross-check for the robustness of our results. From Combined, we find a tight upper limit at a larger $k$,  at $k = 0.383 \, \rm{Mpc}^{-1}$, where we find $[\Omega_{\HI}b_{\HI}]_{\rm UL} =  1.06 \times 10^{-2}$. The upper limits on $[\Omega_{\HI}b_{\HI}]$ are also presented in Tables~\ref{tab:ul_MLE}, ~\ref{tab:ul_GPR} and ~\ref{tab:ul_comb}.

Figure~\ref{fig:pssph} also consolidates the final results from all the previous works with the same observational data. \citetalias{P22} (light-blue circle) and \citetalias{AE23} (orange asterisk) have used the Total and the Cross TGE, respectively. \citealt{Ch21} (\citetalias{Ch21}) estimated the PS in delay space at multiple redshifts from this data. Here, we present their results (black circle) from the redshift $z = 2.19$. While the above works considered foreground avoidance to constrain the $21$-cm PS, \citetalias{AE23b} have used PF (red square) and  GPR (blue square) to remove the foregrounds.
A compilation of the best upper limits derived using this $25$-hour Band $3$ uGMRT data is given in Table~\ref{tab:results_IM}.  We note that the frequency ranges used in these works are not the same, and the results are not directly comparable. \citetalias{P22}, \citetalias{AE23}, \citetalias{AE23b} have used $24.4\,{\rm MHz}$ bandwidth at the central frequency of $432.8\,{\rm MHz}$ whereas, \citetalias{Ch21} and the present work have respectively used $8\,{\rm MHz}$ and $100\,{\rm MHz}$ bandwidth both at the central frequency of $444\,{\rm MHz}$.

\begin{table}
\centering
\caption{A compilation of the best upper limits derived using the $25$-hour Band $3$ uGMRT data considered in this paper.  The overall tightest constraints obtained using this data are marked with boldface.} 
    \begin{tabular}{cccccc}
        \hline
        \hline
        Works & $z$ & k & $\Delta_{\rm UL}^2(k)$ &
        $[\Omega_{\HI}b_{\HI}]_{\rm UL}$  \\
        & & $\rm{Mpc}^{-1}$ & $\rm{mK}^2$& \\
        \hline
        \multirow{4}{*}{\citetalias{Ch21}} & 1.96 & 0.99 & $(58.57)^2$ & $0.09$\\
         & 2.19 & 0.97 & $(61.49)^2$ & $0.11$\\
         & 2.62 & 0.95 & $(60.89)^2$ & $0.12$\\
         & 3.58 & 0.99 & $(105.85)^2$ &$0.24$\\
        \hline
        \citetalias{P22} & 2.28 & 0.35 & $(133.97)^{2}$ & $0.23$\\
        \hline
        \citetalias{AE23} & $2.28$ & 0.80 & $(58.67)^{2}$  & $0.072$ \\
        \hline
        \citetalias{AE23b} &  &  &  & \\        
        PF & \multirow{2}{*}{$2.28$} & $0.25$ & $(18.07)^2$ & $0.036$\\
        GPR &   & $0.30$ & $(24.54)^2$ & $0.045$\\
        \hline
        \rm{\bf Present work} &   &  &  & \\        
        \multirow{2}{*}{PF} & \multirow{6}{*}{$1.9 - 2.6$} & $\mathbf{0.22}$ & $\mathbf{(4.68)^2}$ & $\mathbf{0.010}$\\
         &  & $0.38$ & $(8.48)^2$ & $0.014$\\
        \multirow{2}{*}{GPR} &   & $0.22$ & $(7.56)^2$ & $0.016$\\
         &  & $0.40$ & $(10.90)^2$ & $0.018$\\
        \multirow{2}{*}{Combined} &   & $0.22$ & $(6.32)^2$ & $0.014$\\
         &  & $\mathbf{0.38}$ & $\mathbf{(6.38)^2}$ & $\mathbf{0.011}$\\
        \hline
    \end{tabular}
    \label{tab:results_IM}
\end{table}

The best upper limit from all the earlier works was $\Delta_{\rm UL}^2(k) = (18.07)^2 \, \rm{mK}^2$ at $k = 0.247 \, \rm{Mpc}^{-1}$ which was obtained in \citetalias{AE23b}. The present upper limit $\Delta_{\rm UL}^2(k) = (4.68)^2 \, \rm{mK}^2$ at $k = 0.219 \, \rm{Mpc}^{-1}$ is approximately $15$ times tighter  than our previous work. The upper limit $[\Omega_{\HI}b_{\HI}] < 1.01\times 10^{-2}$ is also $\sim5$ times tighter than our previous limit $2.2\times 10^{-2}$ (\citetalias{AE23b}).
The large bandwidth and the MAD filter are  the two key factors that lead  to these improved upper limits. The four-fold increase in the bandwidth reduces the noise in the power spectrum by a factor of $\approx4$. This also allows us to probe smaller $k_\parallel$ modes, whereby we are able to  reach  $k = 0.219 \, \rm{Mpc}^{-1}$ in the present analysis compared to the previous analysis where the upper limit was quoted at $k = 0.247 \, \rm{Mpc}^{-1}$. This lowers the $\Delta^2(k)$ values and the uncertainties $\sigma$  by a factor of $\approx1.4$. The MAD filter reduces the r.m.s., giving an improvement of $\approx1.7$. We also note that the reduction in the excess variance tightens the constraints by a factor of $\approx1.8$.

\section{Constraining \texorpdfstring{\omb{}}{Omb}}
\label{sec:omb}

In this section, we use all the available $[C_{\ell}(\Delta\nu)]_{\rm res}$ values to put a constraint on the single parameter $[\Omega_{\HI}b_{\HI}]$ without estimating the spherical PS. This approach was introduced in \citetalias{AE23} where $[\Omega_{\HI}b_{\HI}]$ was estimated from the measured $C_{\ell}(\Delta\nu)$ using foreground avoidance. However, we used it in a slightly different form in \citetalias{AE23b} (Section 6) which we follow here.

We obtain $[\Omega_{\HI}b_{\HI}]^2 = 3.18\times 10^{-5} \pm 4.62 \times 10^{-5}$  and $7.37\times 10^{-5} \pm 9.99 \times 10^{-5}$ with PF and GPR, respectively. We observe that the estimated $[\Omega_{\HI}b_{\HI}]^2$ values lie within $0 \pm 1\sigma$ noise fluctuations. Subsequently, we use these values to derive an upper limit on the expected 21-cm signal, $[\Omega_{\HI}b_{\HI}]_{\rm UL} =  1.11 \times 10^{-2}$ and $[\Omega_{\HI}b_{\HI}]_{\rm UL} =  1.65 \times 10^{-2}$ for PF and GPR, respectively. We obtain the tightest limit from Combined where we find $[\Omega_{\HI}b_{\HI}]^2 = 3.10\times 10^{-5} \pm 3.60 \times 10^{-5}$ which yields  $[\Omega_{\HI}b_{\HI}]_{\rm UL} =  1.01 \times 10^{-2}$. These results are quoted in Table~\ref{tab:results}. The upper limits obtained using this independent technique from our earlier works are also noted in that table.

\begin{table}
\centering
\caption{Constraints on \omb{} obtained directly from $[C_{\ell}(\Delta\nu)]_{\rm{res}}$ without referring to the spherical PS.} 
    \begin{tabular}{ccccc}
        \hline
        \hline       
        \rm{Works} & $z$ & \begin{tabular}[c]{@{}c@{}} \ombsq{} \\ $\times 10^{-5}$ \end{tabular} & SNR &  \begin{tabular}[c]{@{}c@{}} $[\Omega_{\HI}b_{\HI}]_{\rm UL}$\\$\times 10^{-2}$ \end{tabular}  \\
        \hline 
        \citetalias{AE23} & 2.28 & $75.1 \pm 147$ & $0.51$ &   $6.08$ \\
        \hline
        \citetalias{AE23b}   & &  &  &    \\        
        \rm{PF}  & \multirow{2}{*}{$2.28$} & $4.41 \pm 21.7$ & $0.20$ &   $2.19$ \\         
        \rm{GPR}  &  & $14.0 \pm 40.9$ & $0.34$ &   $3.09$ \\        
        \hline         
        Present work  & &  &  &    \\        
        PF  & \multirow{3}{*}{$1.9-2.6$} & $3.18 \pm 4.62$ & $0.69$ &   $1.11$ \\
        GPR & & $7.37 \pm 9.99$ & $0.74$ &   $1.65$ \\
        Combined & & $3.10 \pm 3.60$ & $0.86$ &   $1.01$ \\
        
        \hline
    \end{tabular}
    \label{tab:results}
\end{table}

Previously, \cite{Prochaska2005, Noterdaeme09, Not, Zafar} and \cite{Bird17} have measured the \HI{} mass density $\Omega_{\HI}$ in the redshift range $z\sim 1.9-2.6$ using Damped Lyman-$\alpha$ observations. These studies quote  $\Omega_{\HI}$  roughly in the range $\sim 0.4 \times 10^{-3} - 1.0 \times 10^{-3}$. Using a weighted fit of all the $\Omega_{\HI}$ measurements at $z<6$, \cite{Rhee2018} reported  $0.5 \times 10^{-3} \lesssim \Omega_{\HI} \lesssim 0.9 \times 10^{-3}$ for $z\sim1.9 - 2.6$ with a $95\%$ confidence.  Simulations of the post-EoR 21-cm signal \citep{Deb16} predict the bias parameter $b_{\HI}$ to be close to unity at the $z$-range and $k$-scales probed here. These independent studies suggest that our upper limits on  $[\Omega_{\HI}b_{\HI}]$ is around $10$ times larger than the value predicted by these measurements. A wider bandwidth and longer observations from uGMRT can significantly lower the present upper limit and open up the possibility for the detection of the IM signal at these redshifts.

\section{Summary and Conclusions}
\label{sec:conclusion}

\HI{} $21$-cm intensity mapping (IM) has the potential to become a leading tool in observational cosmology. Several works have measured the 21-cm IM signal by using cross-correlation of \HI{} measurements and optical galaxy surveys. However, it is difficult to detect the faint IM signal in auto-correlation due to the many orders of magnitude brighter Galactic and extra-galactic foregrounds. 

In a series of works, we have explored a deep (25 hours) observation of the ELAIS-N1 field using uGMRT with the aim of 21-cm IM from a redshift around $z\sim2$. In these works, we have used the Tapered Gridded Estimator (TGE), which offers several unique features for estimating the 21-cm PS from radio-interferometric measurements. 
In \citetalias{P22} of this series,  we have used the TGE to analyse a $24.4\,\rm{MHz}$ sub-band of the mentioned data at $432.8 \, \rm {MHz}$ ($z=2.28$). In that paper, the signals in the two polarization states  (RR and LL) were combined to estimate the 21-cm PS.  In \citetalias{AE23}, we showed that the foreground level is substantially reduced if the PS is estimated by cross-correlating the RR and LL polarizations rather than combining them. Both works used foreground avoidance, and only a limited region of the $(k_{\perp},k_{\parallel})$ plane was used to constrain the 21-cm signal. In \citetalias{AE23b}, we have introduced a foreground removal technique which allowed us to utilise the full  $(k_{\perp},k_{\parallel})$ plane, and we obtained a tighter upper limit  $[\Omega_{\HI}b_{\HI}] < 2.2 \times 10^{-2}$ at $k = 0.247 \, {\rm Mpc^{-1}}$ as compared to the earlier works. 

The present paper is the fourth in this series. Here, we have considered $100\,{\rm MHz}$ bandwidth data spanning the frequency range $394-494$~MHz $(z=1.9-2.6)$.  The bandwidth under consideration is four times larger than the earlier works (details in Section~\ref{sec:data}). To analyse this wideband data, we have incorporated baseline migration in TGE, which was not done in the earlier works. The (Wideband) TGE is briefly discussed in Section~\ref{sec:wtge} while the detailed formalism of TGE can be found in \citetalias{AE23b} (and references therein). Using the TGE, we estimated the Multi-frequency Angular Power Spectrum (MAPS) \clb{} on each grid point $\ellb$. We then used the two foreground removal tools developed in \citetalias{AE23b}, namely polynomial fitting (PF) and Gaussian Process Regression (GPR) on the estimated \clb{}. 
PF and GPR are briefly discussed in Section~\ref{sec:FGremoval}.
For each of these two methods, we have separately identified 
the $\ellb$ grids where  foreground removal was successful and the remaining $\ellb$ grids were flagged.  We found $43$  $\ellb$ grids where PF worked and $47$ $\ellb$ grids where GPR worked, but only $6$ $\ellb$ grids  where both worked. Figures~\ref{fig:cl} shows the results for a few representative $\ellb$ grids from each of the different categories discussed above.  In addition to PF and GPR, we also considered a Combined data containing $84$ $\ellb$ grids where either PF or GPR or both work. 

We note that our flagging criteria is entirely based on the residual  $[C_{\ellb}(\Delta\nu)]_{\rm res}$ in the range $\Delta \nu \leq [\Delta \nu]$, and it is possible that we include some $\ellb$ grids where the foreground model provides a poor fit to the measured \clb{} in the range $\Delta \nu > [\Delta \nu]$.  The third row of Figure~\ref{fig:cl} is an example of the situation where this occurs. It is however useful to note that the grids that have a poor fit in the range $\Delta \nu > [\Delta \nu]$,  have large predicted error bars for the residuals in the range $\Delta \nu \leq [\Delta \nu]$. The maximum likelihood estimator for the final 21-cm PS  automatically gives a lower weight to the contribution from such grids. An alternative approach would be to visually identify and flag such $\ellb$ grids. We have repeated the entire analysis after manually flagging the particular $\ellb$ grid mentioned above. We find that removing this $\ell$ grid systematically reduces the  upper limits on  \omb{} (Table~\ref{tab:ul_GPR})  by $6 - 9 \%$ across all the $k$ bins. However, we have not incorporated any manual flagging for the results reported here, and the decision to include an $\ellb$  grid or not is entirely set by the objective criteria presented in Section~\ref{sec:FGremoval}.

We have estimated  the cylindrical PS \pkb{} from the  measured \clb{} as well as  the  residual $[C_{\ellb}(\Delta\nu)]_{\rm res}$ after foreground removal. Note that this was carried out individually for each $\ellb$ grid without binning the data. To visualise the results in Figures~\ref{fig:cylps} and \ref{fig:pkslice}, we have binned the values of \clb{} into equally spaced annular bins in the $uv$-plane and shown the binned cylindrical PS \pk{}. In Figure~\ref{fig:cylps} we noticed that the large values $\mid P(k_{\perp}, k_{\parallel}) \mid > 10^5 \, {\rm mK^2}$ are removed by PF and GPR, and the residual $\mid P(k_{\perp}, k_{\parallel}) \mid$ are found to have values in the range $10^3 - 10^5 \, {\rm mK^2}$. 
We showed $\mid P(k_{\perp}, k_{\parallel}) \mid$ as a function of $k_\parallel$ in Figure~\ref{fig:pkslice}. We found that although foregrounds are largely subtracted, the PS $\mid P(k_{\perp}, k_{\parallel}) \mid$ are not entirely consistent with the system noise level. However, the values of $\mid P(k_{\perp}, k_{\parallel}) \mid$ are found to be within the statistical uncertainties when we  combine the system noise with the foreground modelling errors.

We have studied the statistics of the estimated  $P(\kb, k_{\parallel})$ through the quantity $X$ (equation~\ref{eq:xstat}). $X$ is expected to be symmetric around zero with mean $(\mu =0)$ and have a unit standard deviation $(\sigma_{\rm Est} \sim 1)$ if foregrounds are perfectly subtracted . We find $X$ to be symmetric with $\mu\approx 0$, which indicates that foregrounds are largely subtracted.   However, we find an excess variance $(\sigma_{\rm Est} > 1)$, indicating that the actual statistical fluctuations exceed  our error predictions.  The error predictions are scaled up by the factor $\sigma_{\rm Est}$ to account for the excess variance in the data.

We have used $[C_{\ellb}(\Delta\nu)]_{\rm res}$ with the scaled error predictions in a maximum likelihood estimator to obtain the spherical PS $P(k)$ and its uncertainties. We have used these to obtain the mean squared 21-cm brightness temperature fluctuations  $\Delta^2(k)$. In majority of the  cases the estimated values are found to be within $0 \pm 3 \sigma$, although  there are a few cases also where the values exceed $0 \pm 5 \sigma$, possibly due to some remaining low level foreground contamination. 
For  PF, we find the tightest $2\sigma$ upper limits $\Delta_{\rm UL}^2(k) = (4.68)^2 \, \rm{mK}^2$ at $k = 0.219 \, \rm{Mpc}^{-1}$ which yields $[\Omega_{\HI}b_{\HI}]_{\rm UL} =  1.01 \times 10^{-2}$. The upper limits obtained from GPR and Combined are also close to this value, which shows the robustness of our results. 
 The upper limits and other relevant details for PF, GPR and Combined are presented in Tables~\ref{tab:ul_MLE}, ~\ref{tab:ul_GPR} and ~\ref{tab:ul_comb} respectively.

The PS results from the present work and all the earlier works using  this Band $3$ observation are shown together in Figure~\ref{fig:pssph}. The best upper limits from these works are all compiled in Table~\ref{tab:results_IM}. The  upper limit $\Delta_{\rm UL}^2(k) = (4.68)^2 \, \rm{mK}^2$ at $k = 0.219 \, \rm{Mpc}^{-1}$  obtained in the present work is approximately $15$ times tighter than the previous upper limit $\Delta_{\rm UL}^2(k) = (18.07)^2 \, \rm{mK}^2$ at $k = 0.247 \, \rm{Mpc}^{-1}$ (\citetalias{AE23b}). The tightest upper limit  $[\Omega_{\HI}b_{\HI}]_{\rm UL} =  1.01 \times 10^{-2}$ obtained in the present work from the Combined data is found to be  nearly $5$ times improved over the previous limit $[\Omega_{\HI}b_{\HI}]_{\rm UL} =  2.2 \times 10^{-2}$ (\citetalias{AE23b}). 

In the present  work, we have obtained a nearly foreground-free PS from uGMRT wideband data. We have found a tight constraint $[\Omega_{\HI}b_{\HI}] < 1.01 \times 10^{-2}$ on the 21-cm signal, which is within an order of magnitude to the expected signal (Section~\ref{sec:omb}).  Deeper and larger bandwidth observations from uGMRT can significantly lower this limit and open up the possibility for the detection of the IM signal at these redshifts. This wideband IM analysis technique is also promising for the ongoing MeerKAT observations \citep{Mauch2020} and the forthcoming Hydrogen Intensity and Real-time Analysis eXperiment (HIRAX; \citealt{Crichton2022}), which can possibly provide cleaner (RFI-free) wideband data than the data considered here. This analysis is also suitable for CHIME (\citealt{chimeIM}) which covers large angular scales.   On a separate note, the present formalism can also be  adapted to low frequency $(\sim 150 \,{\rm MHz})$ Epoch of Reionization (EoR) observations with the uGMRT, Hydrogen Epoch of Reionization Array (HERA; \citealt{Abdurashidova_2022}), LOw-Frequency ARray (LOFAR; \citealt{mertens20}) and Murchison Widefield Array (MWA; \citealt{Trott2020, Patwa2021}).  We plan to address EoR observations in future work. 

\section*{Acknowledgements}

We thank the anonymous reviewer for a careful reading of the manuscript and for the comments. We thank the staff of GMRT for making this observation possible. GMRT is run by the National Centre for Radio Astrophysics (NCRA) of the Tata Institute of Fundamental Research (TIFR). AG would like to thank IUCAA, Pune, for providing support through the associateship programme.  AG wish to acknowledge funding provided under the SERB-SURE grant SUR/2022/000595 of the Science \& Engineering Research Board, a statutory body of the Department of Science \& Technology (DST), Government of India. Part of this work has used the Supercomputing facility `PARAM Shakti' of IIT Kharagpur established under the National Supercomputing Mission (NSM), Government of India and supported by the Centre for Development of Advanced Computing (CDAC), Pune.

\section*{Data Availability}

The observed data are publicly available through the GMRT data archive\footnote{\url{https://naps.ncra.tifr.res.in/goa/}} under the proposal code $32\_120$. The simulated data used here are available upon reasonable request to the corresponding author.



\bibliographystyle{mnras}
\bibliography{myref} 

\appendix

\section{Choice of covariance functions for GPR}
\label{app:covariance}

In this appendix, we have considered four different covariance functions (kernels) for the GPR analysis  (Section~\ref{sec:FGremoval}). The kernels considered here are all of the form 
\begin{equation}
    k_{\rm{FG}}(\Delta\nu_m, \Delta\nu_n) = c_1 \, (\Delta\nu_m \cdot \Delta\nu_n + b )^{P}  + c_2 \,  \cos\left(\frac{2\,\pi\,|\Delta\nu_m - \Delta\nu_n|}{Q} \right)
    \label{eq:tot_kernel}
\end{equation}
which is the sum of a polynomial kernel of order $P$ and a cosine kernel.  Here, $c_1, c_2, b$, and $Q$ are hyper-parameters. Of the four kernels considered here, two 
(K1. $P=2$ and K2. $P=3$) incorporate only  the polynomial kernel ($c_2=0$), and the other two 
(K3. $P=2+\cos$ and K4. $P=3+\cos$) incorporate both the polynomial and cosine components. 
We have used   \clb{} measured in the range $\Delta\nu > [\Delta\nu]$ to estimate the optimal values of the hyper-parameters of the kernel. We subsequently use the optimized kernel to make foreground model predictions $\left[C_\ell(\Delta\nu)\right]_{\rm FG}$ in the range $\Delta\nu \leq [\Delta\nu]$.  

\begin{figure}
    \centering
    \includegraphics[width=0.9\columnwidth]{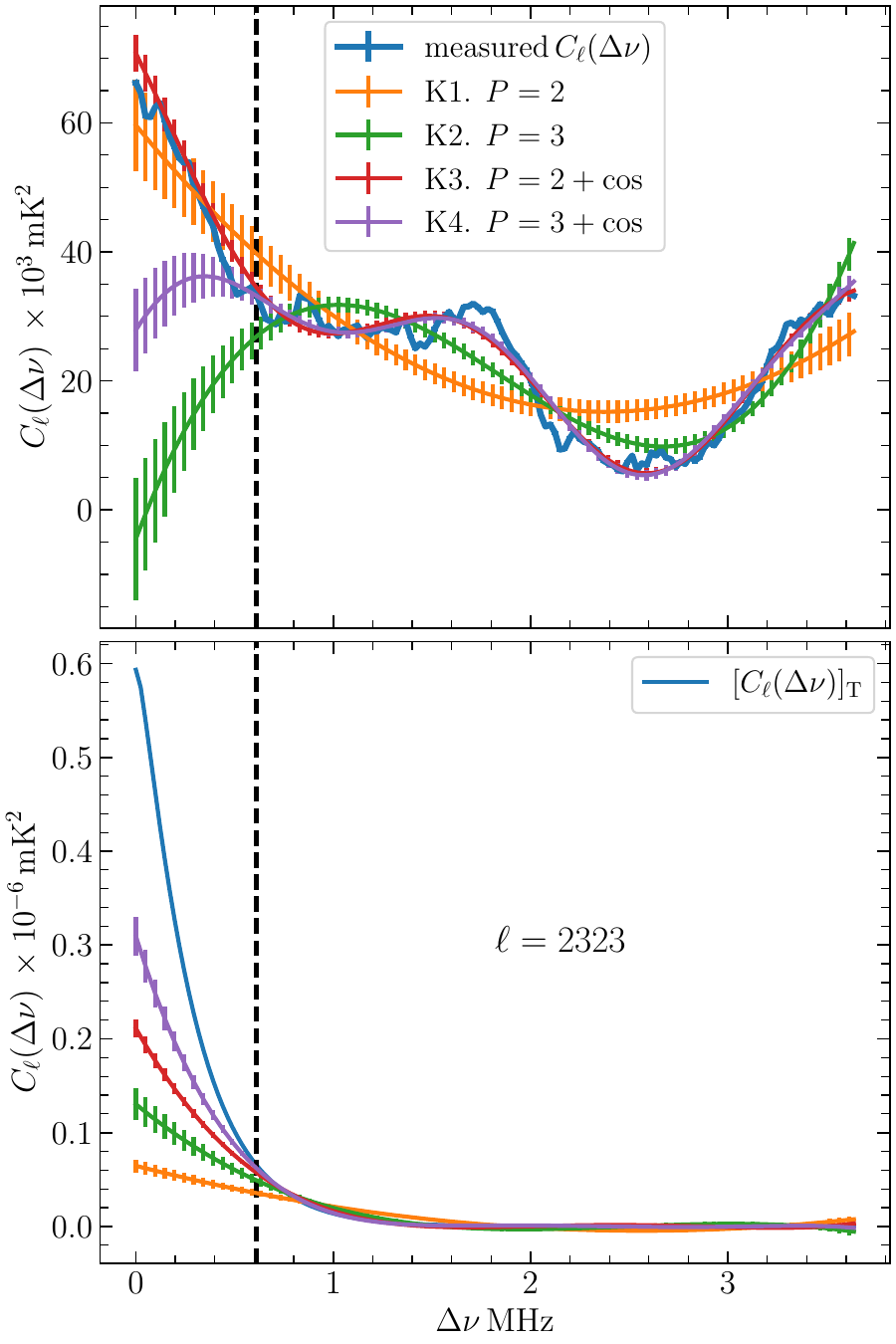}
    \caption{This figure shows the effect of choosing different covariance functions in GPR. A representative $\ellb$ with $\ell = 2323$, previously shown in Figure~\ref{fig:cl}, has been chosen for the demonstration. The top panel shows the measured \clb{} and the foreground model predictions $\left[C_{\ellb}(\Delta\nu)\right]_{\rm FG}$ obtained from different choices of the kernels mentioned in the legend. The error bars show $2\sigma$ uncertainties in the foreground modelling. The bottom panel shows the GPR model predictions when we apply the same kernels on $\left[C_{\ell}(\Delta\nu)\right]_T$ the theoretically predicted 21-cm signal. The bottom panel presents a visual demonstration of how higher-order polynomials, or complicated kernels, give rise to larger signal loss.}
    \label{fig:covariances}
\end{figure}
\begin{figure}
    \centering
    \includegraphics[width=\columnwidth]{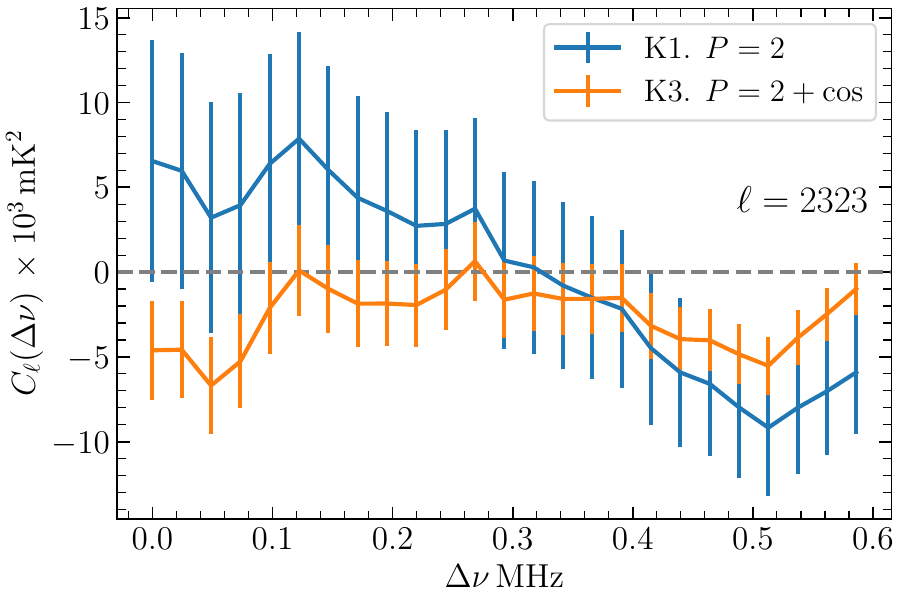}
    \caption{This shows $\left[C_{\ellb}(\Delta\nu)\right]_{\rm res}$ in the range $\Delta \nu \le [\Delta \nu]$ for the two kernels mentioned in the legend. The $2\sigma$ error bars represent the total errors, which combine the errors due to system noise and foreground modelling, and the gray horizontal line shows zero. 
    }
    \label{fig:covariances_res}
\end{figure}

The analysis  presented in this appendix is entirely restricted to  a particular $\ellb$ for which the results have already been shown in the third row of Figure~\ref{fig:cl} of the main text.  The \clb{} measured at this $\ellb$ exhibits  a prominent oscillatory pattern, which is why we have chosen this  for demonstration purpose. The top panel of Figure~\ref{fig:covariances} shows the measured \clb{} along with the  GPR foreground model predictions $\left[C_{\ellb}(\Delta\nu)\right]_{\rm FG}$ and the corresponding  $2\sigma$ uncertainties for the four kernels considered  here. We first consider kernel K1, which is the polynomial kernel that  has been adopted for the entire analysis presented in the main body of the text.  Considering the range $\Delta\nu > [\Delta\nu]$,  we find that the  foreground model $\left[C_{\ellb}(\Delta\nu)\right]_{\rm FG}$  primarily captures the smooth, slowly varying $\Delta\nu$ dependence of the measured \clb{}, and is not much influenced by the oscillatory pattern. Considering the range $\Delta\nu \leq [\Delta\nu]$, we find that the extrapolated values of  $\left[C_{\ellb}(\Delta\nu)\right]_{\rm FG}$ are in reasonably good agreement  with the  measured \clb{}.  We next consider kernel K2, which is a higher order  polynomial kernel as compared to K1. Considering the range $\Delta\nu > [\Delta\nu]$,  we now find that the  foreground model $\left[C_{\ellb}(\Delta\nu)\right]_{\rm FG}$  starts responding more to the oscillatory pattern  in the measured \clb{}. Here, the extrpolated values of  $\left[C_{\ellb}(\Delta\nu)\right]_{\rm FG}$ do not match  the  measured \clb{}  in  the range $\Delta\nu \leq [\Delta\nu]$. We next consider K3  where we have augmented  K1  with an additional cosine kernel.   Considering the range $\Delta\nu > [\Delta\nu]$,  we now find that the foreground model $\left[C_{\ellb}(\Delta\nu)\right]_{\rm FG}$ closely matches the measured \clb{}, including the oscillatory pattern which is missed out by K1.  Considering the range $\Delta\nu \leq [\Delta\nu]$, we find that the extrapolated values of  $\left[C_{\ellb}(\Delta\nu)\right]_{\rm FG}$ are in good agreement with the measured \clb{}, with a match that visually appears to be better than that provided by K1. Considering K4, we find that 
$\left[C_{\ellb}(\Delta\nu)\right]_{\rm FG}$ closely matches K3 in the range $\Delta\nu > [\Delta\nu]$, however it goes completely off in the range  $\Delta\nu \leq [\Delta\nu]$ where the match to the measured \clb{} is the worst  among the four kernels considered here. Finally, we see that the kernels K2  and K4 fail to match the measured \clb{} in the range  $\Delta\nu \leq [\Delta\nu]$, and we cannot use these for foreground subtraction for this particular $\ellb$ grid. In summary, kernels K1 and K3 work well to predict 
$\left[C_{\ellb}(\Delta\nu)\right]_{\rm FG}$ in the range  $\Delta\nu \leq [\Delta\nu]$.

The bottom panel of Figure~\ref{fig:covariances} shows the expected 21-cm signal   $\left[C_{\ell}(\Delta\nu)\right]_T$.    In order to assess the  21-cm signal loss, we have applied foreground subtraction to  $\left[C_{\ell}(\Delta\nu)\right]_T$.   Considering the range $\Delta\nu > [\Delta\nu]$,  we find that \clb{} is close to zero for $\Delta \nu > 1.2 \, {\rm MHz}$.  The behaviours of  the  foreground model $\left[C_{\ellb}(\Delta\nu)\right]_{\rm FG}$ is primarily decided by the values of \clb{} in the relatively small range $[\Delta \nu] < \Delta \nu \le  1.2 \, {\rm MHz} $. We see that the smooth, slowly varying polynomial kernel K1 is not very sensitive to this, and the foreground model prediction  $\left[C_{\ellb}(\Delta\nu)\right]_{\rm FG}$  is close to zero in the range $\Delta\nu \leq [\Delta\nu]$. The foreground model predictions  in the range  $\Delta\nu \leq [\Delta\nu]$  increase successively as we change the kernel from K1 to K2, K3, and K4. We also expect the 21-cm signal loss due to foreground subtraction to increase in the same way if we vary the kernel from K1 to K4. In order to quantify the 21-cm signal loss, we have applied foreground subtraction to $\left[C_{\ell}(\Delta\nu)\right]_T$ considering all the available $\ellb$ grids, and used the residuals to estimate the binned 21-cm PS. We have compared the recovered  21-cm PS  with the input model to estimate the percentage of 21-cm signal loss.  Table~\ref{tab:signal_loss} presents the results for the four different kernels considered here.  We have the least 21-cm signal loss for K1 where it is $50 \%$ for the lowest $k$ bin $(0.2 \, {\rm Mpc}^{-1}$), and less than $25 \%$ for all the other $k$ bins.  
The higher-order polynomial kernel K2 ranks second, and the 21-cm signal loss increases considerably for kernels K3 and K4 which incorporate an additional cosine kernel.  In summary, we have the least 21-cm signal loss for  the $P=2$ polynomial kernel, and the loss increases if we increase the order of the polynomial kernel or include an additional cosine kernel. 

\begin{table}
    \centering
    \caption{This presents the signal loss (in percentage) from different covariance functions for GPR.}
    \begin{tabular}{ccccc}
    \hline
    \hline
    k & {\rm K1.} $P=2$  & {\rm K2.} $P=3$  & {\rm K3.} $P=2$ & {\rm K4.} $P=3$\\
    $\rm{Mpc}^{-1}$ &  &   & $+ \cos$ & $+ \cos$ \\
    \hline 
    0.2 & 50.0  & 69.7 & 84.3 & 97.0 \\
    0.4 & 20.2  & 37.0 & 52.1 & 73.8 \\
    0.6 & 9.80  & 26.6 & 40.2 & 67.9 \\
    1.0 & 1.25  & 8.85 & 16.9 & 51.3 \\
    2.0 & 10.4  & 23.9 & 37.5 & 75.5 \\
    3.5 & 1.44  & 31.0 & 49.2 & 98.4 \\
    6.2 & 16.4  & 56.4 & 82.2 & 27.9 \\
    \hline
    \end{tabular}
    \label{tab:signal_loss}
\end{table}


\begin{table}
    \centering
    \caption{This shows the measured $\Delta^2(k)$, corresponding errors $\sigma(k)$, ${\rm SNR} = \Delta^2(k)/\sigma(k)$, and the $2\,\sigma$ upper limits   $\Delta_{\rm UL}^{2}(k)$ and  $[\Omega_{\HI}b_{\HI}]_{\rm UL}$ for GPR when we use the kernel K3, which combines a $P=2$ polynomial kernel and a cosine kernel.}
        \begin{tabular}{cccccc}
        \hline
        \hline
        $k$ & $\Delta^2(k)$ & $1\sigma(k)$ & SNR & $\Delta_{\rm UL}^{2}(k)$ & $[\Omega_{\HI}b_{\HI}]_{\rm UL}$ \\
        $\rm{Mpc}^{-1}$ & $\rm{mK}^2$ & $\rm{mK}^2$ & & $\rm{mK}^2$  & $\times 10^{-2}$\\
        \hline

        $0.198$ &  $-(5.08)^2$ & $(6.41)^2$ & $-0.63$ & $(9.06)^2$ & $2.09$ \\
        $0.351$ &  $-(22.38)^2$ & $(7.61)^2$ & $-8.64$ & $(10.77)^2$ & $1.86$ \\
        $0.623$ &  $(37.14)^2$ & $(11.85)^2$ & $9.82$ & $(40.75)^2$ & $5.56$ \\
        $1.106$ &  $(89.66)^2$ & $(23.55)^2$ & $14.50$ & $(95.64)^2$ & $10.67$ \\
        $1.962$ &  $(48.64)^2$ & $(57.21)^2$ & $0.72$ & $(94.40)^2$ & $8.84$ \\
        $3.481$ &  $(290.12)^2$ & $(127.42)^2$ & $5.18$ & $(341.53)^2$ & $27.49$ \\
        $6.176$ &  $-(544.26)^2$ & $(458.95)^2$ & $-1.41$ & $(649.05)^2$ & $45.74$ \\

        \hline
    \end{tabular}
    \label{tab:ul_MLE_cos}
\end{table}

We now focus our attention on the two kernels K1 and K3,  for which 
$\left[C_{\ellb}(\Delta\nu)\right]_{\rm FG}$  reasonably matched the \clb{} measured in the range $\Delta\nu \leq [\Delta\nu]$. The residuals after foreground subtraction are presented in Figure~\ref{fig:covariances_res}. For both the kernels, the residuals are an order of magnitude smaller than the measured \clb{}, and are comparable to the predicted error bars which are dominated by the uncertainties in the GPR predictions. For the present analysis, we have 
used the quantity ${\bf A}/ {\bf dA} $  (defined through equation~\ref{eq:cl_Pk_sinc_p4}) to quantify the amplitude of the residuals.   The criteria  $\abyda > 3$ has been used to identify the grids where foreground subtraction fails,  and these grids are flagged out from the subsequent analysis.  For the particular $\ellb$ being analyzed here, we obtain  ${\bf A/dA} = 2.86$ and $ -9.16$ for kernels K1 and K3 respectively. This implies that the present $\ellb$ would be accepted for the subsequent analysis if we use kernel K1, whereas it would be flagged out if we use kernel K3.  Note the larger error predictions for kernel K1 as compared to those for kernel K3. The fact that kernel K1 is unable to model the oscillatory patterns present in the measured \clb{} at $\Delta\nu > [\Delta\nu]$, is reflected as large error bars for the foreground predictions in the range $\Delta\nu \leq [\Delta\nu]$, and these large error bars are also propagated to $\left[C_{\ellb}(\Delta\nu)\right]_{\rm res}$. 

Although foreground subtraction fails for this particular $\ellb$ if we use kernel K3, there are several other $\ellb$ where this kernel successfully subtracts the foregrounds. For a more comprehensive comparison, we have used kernel K3 on the entire grid, and the results are presented in  Table~\ref{tab:ul_MLE_cos}.  Considering  $\Delta^{2}(k)$, we see that the values are not consistent with $ 0 \pm 2 \sigma $ for any of the $k$-bins. Further,  the first two $k$-bins  have negative values of  $\Delta^{2}(k)$ that are in excess of the $2 \sigma$ fluctuations, hinting at the possibility that using kernel K3 may be overfitting and introducing some negative systematics. Comparing with the results for kernel K1  (shown in Table~\ref{tab:ul_GPR}), we find that values of $\Delta_{\rm UL}^{2}(k)$ are comparable to those for kernel K1 at the second and third $k$-bins, whereas they are $\sim 1.5 - 3.6$ times larger at the other bins.  

Based on the relative performance of the various kernels considered here, we have decided to use kernel K1 for the results presented in the main text of this paper. Kernel K1 has the advantage that it is the simplest of the four kernels that we have considered. It only captures the smooth, slowly varying $\Delta \nu$ variation and is reasonably successful in modelling the behavior of the measured \clb{} in the range $\Delta\nu \le [\Delta\nu]$. It has the added advantage that it leads to the minimum loss of the 21-cm signal. However, it fails to capture the oscillations seen in the measured \clb{} at  $\Delta\nu > [\Delta\nu]$. It is not clear at present whether it is desirable to model these oscillations or not, as kernel K3 which models these oscillations, does not perform any better. The fact that kernel K1 fails to model the oscillations is reflected in the foreground modelling error predictions in the range $\Delta\nu \le [\Delta\nu]$. This ensures that the final results, which combine all the grids, receive a relatively smaller contribution from the $\ellb$  that have large unmodelled oscillations as compared to that from the $\ell$  that have small unmodelled oscillatory patterns.

\section{Results with different flagging criteria}
\label{app:flag}

Table~\ref{tab:diff_flag} shows the results we obtain using different flagging criteria set by $\abyda = 2,3,4\,{\rm and}\, 5$ respectively. We find that for both PF and GPR, the results are very similar for the cases with $\abyda > 2$ and $\abyda > 3$. For $\abyda > 4$, however, we find a negative value at the first $k$-bin for GPR, which is not consistent with zero. It implies that we have included some $\ellb$ where GPR has over-predicted the foregrounds, leaving large negative residuals. We also find that for the criteria $\abyda > 5$, there are large positive values at different $k$ bins, which are not consistent with noise for both PF and GPR. 
Based on these considerations, we have adopted the flagging criteria $\abyda = 4$  and  $\abyda =   3$ for PF and GPR, respectively.


\begin{landscape}
\begin{table}
    \begin{tabular}{c|cccccc|cccccc}
    \hline
    \hline
    \multirow{3}{*}{$\abyda$} & \multicolumn{6}{c|}{PF} & \multicolumn{6}{c}{GPR} \\
    & $k$ & $\Delta^2(k)$ & $1\sigma(k)$ & SNR & $\Delta_{\rm UL}^{2}(k)$ & $[\Omega_{\HI}b_{\HI}]_{\rm UL}$
    & $k$ & $\Delta^2(k)$ & $1\sigma(k)$ & SNR & $\Delta_{\rm UL}^{2}(k)$ & $[\Omega_{\HI}b_{\HI}]_{\rm UL}$ \\

    & $\rm{Mpc}^{-1}$ & $\rm{mK}^2$ & $\rm{mK}^2$ & & $\rm{mK}^2$  & $\times 10^{-2}$ 
    & $\rm{Mpc}^{-1}$ & $\rm{mK}^2$ & $\rm{mK}^2$ & & $\rm{mK}^2$  & $\times 10^{-2}$ \\
    \hline 
   
    \multirow{7}{*}{2} &  $0.219$ &  $-(6.25)^2$ & $(4.17)^2$ & $-2.25$ & $(5.89)^2$ & $1.28$ & $0.233$ &  $-(6.21)^2$ & $(5.62)^2$ & $-1.22$ & $(7.95)^2$ & $1.68$ \\

    & $0.384$ &  $(8.16)^2$ & $(6.22)^2$ & $1.72$ & $(12.00)^2$ & $2.00$ & $0.404$ &  $-(6.17)^2$ & $(8.27)^2$ & $-0.56$ & $(11.69)^2$ & $1.90$ \\
    
    & $0.671$ &  $(6.04)^2$ & $(10.11)^2$ & $0.36$ & $(15.52)^2$ & $2.06$ & $0.702$ &  $(17.20)^2$ & $(13.82)^2$ & $1.55$ & $(26.03)^2$ & $3.40$ \\
    
    & $1.175$ &  $(53.78)^2$ & $(22.28)^2$ & $5.83$ & $(62.33)^2$ & $6.82$ & $1.219$ &  $(51.71)^2$ & $(29.78)^2$ & $3.02$ & $(66.69)^2$ & $7.21$ \\
    
    & $2.057$ &  $-(5.37)^2$ & $(46.19)^2$ & $-0.01$ & $(65.33)^2$ & $6.04$ & $2.116$ &  $(31.30)^2$ & $(59.62)^2$ & $0.28$ & $(89.94)^2$ & $8.25$ \\
    
    & $3.600$ &  $(249.93)^2$ & $(89.29)^2$ & $7.83$ & $(280.02)^2$ & $22.35$ & $3.674$ &  $(212.43)^2$ & $(121.27)^2$ & $3.07$ & $(273.02)^2$ & $21.69$ \\

    & $6.302$ &  $-(303.53)^2$ & $(183.67)^2$ & $-2.73$ & $(259.75)^2$ & $18.22$ & $6.380$ &  $-(294.80)^2$ & $(253.46)^2$ & $-1.35$ & $(358.44)^2$ & $25.08$ \\

    \hline

    \multirow{7}{*}{3} & $0.219$ &  $-(3.40)^2$ & $(3.61)^2$ & $-0.89$ & $(5.10)^2$ & $1.11$ & $0.233$ &  $(2.95)^2$ & $(4.92)^2$ & $0.36$ & $(7.56)^2$ & $1.60$ \\

    & $0.384$ &  $(8.18)^2$ & $(5.26)^2$ & $2.42$ & $(11.05)^2$ & $1.84$ & $0.404$ &  $-(7.96)^2$ & $(7.71)^2$ & $-1.07$ & $(10.90)^2$ & $1.77$ \\
    
    & $0.671$ &  $(5.30)^2$ & $(8.55)^2$ & $0.38$ & $(13.20)^2$ & $1.75$ & $0.702$ &  $(25.23)^2$ & $(12.39)^2$ & $4.15$ & $(30.72)^2$ & $4.01$ \\
    
    & $1.175$ &  $(23.47)^2$ & $(18.99)^2$ & $1.53$ & $(35.67)^2$ & $3.90$ & $1.219$ &  $(65.18)^2$ & $(26.87)^2$ & $5.88$ & $(75.45)^2$ & $8.15$ \\
    
    & $2.057$ &  $(66.05)^2$ & $(39.00)^2$ & $2.87$ & $(86.04)^2$ & $7.95$ & $2.116$ &  $(59.79)^2$ & $(53.88)^2$ & $1.23$ & $(96.86)^2$ & $8.88$ \\
    
    & $3.601$ &  $(171.56)^2$ & $(75.54)^2$ & $5.16$ & $(202.11)^2$ & $16.13$ & $3.675$ &  $(171.36)^2$ & $(108.16)^2$ & $2.51$ & $(229.70)^2$ & $18.25$ \\
    
    & $6.303$ &  $-(250.75)^2$ & $(155.58)^2$ & $-2.60$ & $(220.02)^2$ & $15.44$ & $6.380$ &  $-(357.48)^2$ & $(228.08)^2$ & $-2.46$ & $(322.55)^2$ & $22.57$ \\
    \hline

    \multirow{7}{*}{4} & $0.219$ &  $-(4.27)^2$ & $(3.31)^2$ & $-1.67$ & $(4.68)^2$ & $1.01$ & $0.219$ &  $-(7.42)^2$ & $(4.01)^2$ & $-3.43$ & $(5.66)^2$ & $1.23$ \\
    
    & $0.384$ &  $(4.28)^2$ & $(5.18)^2$ & $0.68$ & $(8.48)^2$ & $1.41$ & $0.384$ &  $-(7.00)^2$ & $(7.18)^2$ & $-0.95$ & $(10.16)^2$ & $1.69$ \\
    
    & $0.671$ &  $(11.50)^2$ & $(8.34)^2$ & $1.90$ & $(16.47)^2$ & $2.19$ & $0.671$ &  $(32.19)^2$ & $(10.74)^2$ & $8.98$ & $(35.59)^2$ & $4.73$ \\
    
    & $1.175$ &  $(34.12)^2$ & $(18.31)^2$ & $3.47$ & $(42.84)^2$ & $4.68$ & $1.175$ &  $(64.26)^2$ & $(23.44)^2$ & $7.52$ & $(72.31)^2$ & $7.91$ \\

    & $2.057$ &  $(51.56)^2$ & $(37.95)^2$ & $1.85$ & $(74.42)^2$ & $6.88$ & $2.057$ &  $(34.36)^2$ & $(50.17)^2$ & $0.47$ & $(78.83)^2$ & $7.29$ \\
    
    & $3.601$ &  $(168.81)^2$ & $(73.63)^2$ & $5.26$ & $(198.35)^2$ & $15.83$ & $3.601$ &  $(176.12)^2$ & $(100.30)^2$ & $3.08$ & $(226.13)^2$ & $18.05$ \\
    
    & $6.303$ &  $-(249.54)^2$ & $(151.76)^2$ & $-2.70$ & $(214.62)^2$ & $15.06$ & $6.303$ &  $-(319.73)^2$ & $(211.16)^2$ & $-2.29$ & $(298.63)^2$ & $20.95$ \\
    \hline

    \multirow{7}{*}{5} & $0.219$ &  $-(4.05)^2$ & $(3.10)^2$ & $-1.71$ & $(4.38)^2$ & $0.95$ & $0.219$ &  $-(5.80)^2$ & $(3.87)^2$ & $-2.24$ & $(5.48)^2$ & $1.19$ \\
    
    & $0.384$ &  $(7.69)^2$ & $(5.09)^2$ & $2.28$ & $(10.53)^2$ & $1.75$ & $0.384$ &  $-(10.43)^2$ & $(6.89)^2$ & $-2.29$ & $(9.75)^2$ & $1.62$ \\
    
    & $0.671$ &  $-(10.61)^2$ & $(8.08)^2$ & $-1.73$ & $(11.42)^2$ & $1.52$ & $0.671$ &  $(33.97)^2$ & $(10.31)^2$ & $10.87$ & $(36.96)^2$ & $4.91$ \\
    
    & $1.175$ &  $(40.50)^2$ & $(17.72)^2$ & $5.22$ & $(47.63)^2$ & $5.21$ & $1.175$ &  $(48.88)^2$ & $(22.55)^2$ & $4.70$ & $(58.36)^2$ & $6.38$ \\
    
    & $2.057$ &  $(68.01)^2$ & $(36.74)^2$ & $3.43$ & $(85.59)^2$ & $7.91$ & $2.057$ &  $-(56.09)^2$ & $(48.13)^2$ & $-1.36$ & $(68.07)^2$ & $6.29$ \\
    
    & $3.601$ &  $(167.14)^2$ & $(71.33)^2$ & $5.49$ & $(195.22)^2$ & $15.58$ & $3.601$ &  $(171.82)^2$ & $(96.14)^2$ & $3.19$ & $(219.11)^2$ & $17.49$ \\
    
    & $6.303$ &  $-(240.18)^2$ & $(147.31)^2$ & $-2.66$ & $(208.33)^2$ & $14.62$ & $6.303$ &  $-(306.08)^2$ & $(201.53)^2$ & $-2.31$ & $(285.01)^2$ & $20.00$ \\
    \hline

    \end{tabular}
    \caption{This shows the measured $\Delta^2(k)$, corresponding errors $\sigma(k)$, ${\rm SNR} = \Delta^2(k)/\sigma(k)$, and the $2\,\sigma$ upper limits   $\Delta_{\rm UL}^{2}(k)$ and  $[\Omega_{\HI}b_{\HI}]_{\rm UL}$ for PF and GPR considering four different flagging criteria set by $\abyda = 2,3,4\,{\rm and}\, 5$.  }
    \label{tab:diff_flag}
\end{table}
\end{landscape}

\bsp	
\label{lastpage}
\end{document}